\documentclass[pre,twocolumn,showpacs,preprintnumbers,amsmath,amssymb,longbibliography,superscriptaddress]{revtex4-1}
\usepackage[english]{babel}
\usepackage{amssymb,amsmath}
\usepackage{float}
\usepackage{graphicx}
\usepackage{psfrag}
\usepackage{color}
\usepackage{dcolumn}
\usepackage{bm}
\usepackage[normalem]{ulem}
\usepackage[absolute]{textpos}

\newcommand{\new}{\textcolor{red}}
\definecolor{cadetgrey}{rgb}{0.57, 0.64, 0.69}
\newcommand{\old}[1]{{\color{cadetgrey}\sout{#1}}}

\def\beq{\begin{equation}}
\def\eeq{\end{equation}}
\def\bea{\begin{eqnarray}}
\def\eea{\end{eqnarray}}
\newcommand{\stkout}[1]{\ifmmode\text{\sout{\ensuremath{#1}}}\else\sout{#1}\fi}
\usepackage{hyperref}
\hypersetup{
  colorlinks=true,
  citecolor=magenta,
  linkcolor=cyan,
}

\newcommand{\ef}[1]{\textcolor{magenta}{[EF: #1]}}

\begin{document}
\title{Availability versus carrying capacity: Phases of asymmetric exclusion processes competing for finite pools of resources}
\author{Astik Haldar}\email{astik.haldar@gmail.com}
\affiliation{Department of Theoretical Physics \& Center for Biophysics, Saarland University, 66123 Saarbr\"ucken, Germany}
\author{Parna Roy}
\email{parna.roy14@gmail.com}
\affiliation{Shahid Matangini Hazra Government College for Women, Purba Medinipore 721649, West Bengal, India}
\author{Erwin Frey}\email{frey@lmu.de}
\affiliation{Arnold Sommerfeld Center for Theoretical Physics and Center for NanoScience, Department of Physics, Ludwig-Maximilians-Universit\"at M\"unchen, Theresienstra\ss e 37, D-80333 Munich, Germany}
\author{Abhik Basu}\email[corresponding author: ]{abhik.123@gmail.com}
\affiliation{Theory Division, Saha Institute of Nuclear Physics, 1/AF Bidhannagar, Calcutta 700 064, West Bengal, India}

\begin{abstract}
 We address how the interplay between the finite \textit{availability} and \textit{carrying capacity} of particles at different parts of a 
spatially extended system can control the steady state currents and density profiles in the one-dimensional current-carrying 
lanes connecting the different parts of the system. To study this, we set up a minimal model consisting of two particle reservoirs of the same
finite carrying capacity connected by two equally sized anti-parallel asymmetric exclusion processes (TASEP). We focus on the steady-state currents and particle density profiles in the two TASEP lanes. The ensuing phases and the phase diagrams, which can be remarkably complex, are parametrized by the model 
parameters defining particle exchange between the TASEP lanes and the reservoirs and the filling fraction of the particles that determine the total 
resources available.    These parameters may be tuned to make the densities of the two TASEP lanes globally uniform or piece-wise continuous in the form 
of a combination of a single localized domain wall and a spatially 
constant density or a pair of delocalized domain walls.  Our model reveals that the two reservoirs can be preferentially populated or depopulated in the steady states.
\end{abstract}
 
@article{driven-diff1,
  title = {Phase transitions in stationary nonequilibrium states of model lattice systems},
  author = {Katz, S. and Lebowitz, J. L. and Spohn, H.},
  journal = {Phys. Rev. B},
  volume = {28},
  issue = {3},
  pages = {1655-1658},
  numpages = {0},
  year = {1983},
  month = {Aug},
  publisher = {American Physical Society},
  doi = {10.1103/PhysRevB.28.1655},
  url = {https://link.aps.org/doi/10.1103/PhysRevB.28.1655}
}
@article{driven-diff2,
 author= { Katz, S. and Lebowitz, J. L. and Spohn, H.},
 year= {1984},
 month = {Feb},
 title= {Nonequilibrium steady states of stochastic lattice gas models of fast ionic conductors},
 journal= {J. Stat. Phys.},
 pages= {497-537},
 vol= {34},
 url= {https://doi.org/10.1007/BF01018556},
 doi = {10.1007/BF01018556}
}
@book{driven-diff3, 
 title= {Phase transitions and critical phenomena},
 author= { Schmittmann, B. and Zia, R. K.-P. },
 editor= { Lebowitz, Joel Louis and Domb,  Cyril},
 year= {1995},
 vol= {15},
 Publisher= {Academic Press, London}
 }
@book{driven-diff4,
 place={Cambridge},
 title={Nonequilibrium Statistical Mechanics in One Dimension},
 DOI={10.1017/CBO9780511564284},
 publisher={Cambridge University Press},
 year={1997}
 }
 
 @article{macdonald,
author = {MacDonald, C. T. and Gibbs, J. H. and Pipkin, A. C.},
title = {Kinetics of biopolymerization on nucleic acid templates},
journal = {Biopolymers},
volume = {6},
number = {1},
pages = {1-25},
doi = {https://doi.org/10.1002/bip.1968.360060102},
url = {https://onlinelibrary.wiley.com/doi/abs/10.1002/bip.1968.360060102},
eprint = {https://onlinelibrary.wiley.com/doi/pdf/10.1002/bip.1968.360060102},
year = {1968}
}

@article{nat,
 title= {Ribosome recycling, diffusion, and mRNA loop formation in translational regulation},
 vol= {85},
 year= {2003}, 
 month= {Aug},
 page= {755-73},
 author= { Chou T.},
 journal= {Biophys J.},
 doi= {10.1016/S0006-3495(03)74518-4}
 }
 
 @article{tasep-exact-1,
 author= { Derrida, B. and  Domany, E. and Mukamel, D.},
 year= {1992},
 title= {An exact solution of a one-dimensional asymmetric exclusion model with open boundaries},
 journal= {J Stat Phys},
 pages= {667-687},
 vol= {69},
 issue= {3},
 url=  {https://doi.org/10.1007/BF01050430},
 doi= {10.1007/BF01050430}
 }
 @article{tasep-exact-2,
 author= {Sch\"utz, G. and  Domany, E.},
 title= { Phase transitions in an exactly soluble one-dimensional exclusion process},
 journal= {J Stat Phys},
 pages= {277-296},
 vol= {72},
 issue= {1},
 year= {1993},
 url= {https://doi.org/10.1007/BF01048050},
 doi= {10.1007/BF01048050}
} 
@article{tasep-exact-3,
  title = {Time-dependent correlation functions in a one-dimensional asymmetric exclusion process},
  author = {Sch\"utz, G.},
  journal = {Phys. Rev. E},
  volume = {47},
  issue = {6},
  pages = {4265--4277},
  numpages = {0},
  year = {1993},
  month = {Jun},
  publisher = {American Physical Society},
  doi = {10.1103/PhysRevE.47.4265},
  url = {https://link.aps.org/doi/10.1103/PhysRevE.47.4265}
}
@article{krug,
  title = {Boundary-induced phase transitions in driven diffusive systems},
  author = {Krug, J.},
  journal = {Phys. Rev. Lett.},
  volume = {67},
  issue = {14},
  pages = {1882--1885},
  numpages = {0},
  year = {1991},
  month = {Sep},
  publisher = {American Physical Society},
  doi = {10.1103/PhysRevLett.67.1882},
  url = {https://link.aps.org/doi/10.1103/PhysRevLett.67.1882}
}
@article{krug1,
 author= { Derrida, B. and  Janowsky, S. A. and Lebowitz, J. L. and Speer, E. R. },
 title= { Exact solution of the totally asymmetric simple exclusion process: Shock profiles},
 journal= { J Stat Phys},
 pages=  {813--842},
 vol= { 73},
 isuue= {5},
 url= {https://doi.org/10.1007/BF01052811},
 doi= { 10.1007/BF01052811}
 }
 @article{krug2,
  title={Exact correlation functions in an asymmetric exclusion model with open boundaries},
  author={Derrida, B. and Evans, M. R},
  journal={Journal de Physique I},
  volume={3},
  number={2},
  pages={311--322},
  year={1993},
  publisher={EDP Sciences},
  doi= {https://doi.org/10.1051/jp1:1993132 }
}

@article{derrida,
	author = {Derrida, B. and  Evans, M. R. and Hakim, V.  and Pasquier, V. },
	title = {Exact solution of a 1D asymmetric exclusion model using a matrix formulation},
	journal = {J. Phys. A: Math. and Gen.},
    doi = {10.1088/0305-4470/26/7/011},
	url = {https://doi.org/10.1088/0305-4470/26/7/011},
	year = 1993,
	month = {apr},
	publisher = {{IOP} Publishing},
	volume = {26},
	number = {7},
	pages = {1493--1517}
 }
 @article{protein-1,
  title = {Totally asymmetric exclusion process with extended objects: A model for protein synthesis},
  author = {Shaw, Leah B. and Zia, R. K. P. and Lee, Kelvin H.},
  journal = {Phys. Rev. E},
  volume = {68},
  issue = {2},
  pages = {021910},
  numpages = {17},
  year = {2003},
  month = {Aug},
  publisher = {American Physical Society},
  doi = {10.1103/PhysRevE.68.021910},
  url = {https://link.aps.org/doi/10.1103/PhysRevE.68.021910}
}
@article{protein-2,
  title = {Clustered Bottlenecks in mRNA Translation and Protein Synthesis},
  author = {Chou, T. and Lakatos, G.},
  journal = {Phys. Rev. Lett.},
  volume = {93},
  issue = {19},
  pages = {198101},
  numpages = {4},
  year = {2004},
  month = {Nov},
  publisher = {American Physical Society},
  doi = {10.1103/PhysRevLett.93.198101},
  url = {https://link.aps.org/doi/10.1103/PhysRevLett.93.198101}
}
@article{ protein-3,
 author= { Dong, J. J. and Schmittmann, B. and  Zia, R. K. P.},
 year= { 2007},
 title= { Towards a Model for Protein Production Rates},
 journal= {J Stat Phys},
 pages= {21--34},
 vol= {128},
 issue= {1},
 url= { https://doi.org/10.1007/s10955-006-9134-7},
 doi= { 10.1007/s10955-006-9134-7}
} 

@article{PROTEIN-4,
  title = {Inhomogeneous exclusion processes with extended objects: The effect of defect locations},
  author = {Dong, J. J. and Schmittmann, B. and Zia, R. K. P.},
  journal = {Phys. Rev. E},
  volume = {76},
  issue = {5},
  pages = {051113},
  numpages = {13},
  year = {2007},
  month = {Nov},
  publisher = {American Physical Society},
  doi = {10.1103/PhysRevE.76.051113},
  url = {https://link.aps.org/doi/10.1103/PhysRevE.76.051113}
}
@article{traffic-1,
  title={Vehicular traffic: A system of interacting particles driven far from equilibrium},
  author={Chowdhury, D. and Santen, L. and Schadschneider, A.},
  journal={Current Science},
  pages={411--419},
  year={1999},
  publisher={JSTOR}
}
@article{traffic-2,
	doi = {10.1088/0305-4470/34/6/103},
	url = {https://doi.org/10.1088/0305-4470/34/6/103},
	year = {2001},
	month = {feb},
	publisher = {{IOP} Publishing},
	volume = {34},
	number = {6},
	pages = {L45--L52},
	author = {V Popkov and L Santen and A Schadschneider and G M Schütz},
	title = {Empirical evidence for a boundary-induced nonequilibrium phase transition},
	journal = {J. Phys. A: Math. and Gen.}
}
@article{traffic-3,
title = {Statistical physics of vehicular traffic and some related systems},
author = {D. Chowdhury and L. Santen and A. Schadschneider},
journal = {Phys. Rep.},
volume = {329},
number = {4},
pages = {199-329},
year = {2000},
issn = {0370-1573},
doi = {https://doi.org/10.1016/S0370-1573(99)00117-9},
url = {https://www.sciencedirect.com/science/article/pii/S0370157399001179}
}
@article{traffic-4,
  title = {Traffic and related self-driven many-particle systems},
  author = {Helbing, D.},
  journal = {Rev. Mod. Phys.},
  volume = {73},
  issue = {4},
  pages = {1067--1141},
  numpages = {0},
  year = {2001},
  month = {Dec},
  publisher = {American Physical Society},
  doi = {10.1103/RevModPhys.73.1067},
  url = {https://link.aps.org/doi/10.1103/RevModPhys.73.1067}
}
@article{surface-1,
  title = {Dynamic Scaling of Growing Interfaces},
  author = {Kardar, M. and Parisi, G. and Zhang, Yi-Cheng},
  journal = {Phys. Rev. Lett.},
  volume = {56},
  issue = {9},
  pages = {889--892},
  numpages = {0},
  year = {1986},
  month = {Mar},
  publisher = {American Physical Society},
  doi = {10.1103/PhysRevLett.56.889},
  url = {https://link.aps.org/doi/10.1103/PhysRevLett.56.889}
}
@article{surface-2,
  title = {Inhomogeneous growth processes},
  author = {Wolf, D. E. and Tang, Lei-Han},
  journal = {Phys. Rev. Lett.},
  volume = {65},
  issue = {13},
  pages = {1591--1594},
  numpages = {0},
  year = {1990},
  month = {Sep},
  publisher = {American Physical Society},
  doi = {10.1103/PhysRevLett.65.1591},
  url = {https://link.aps.org/doi/10.1103/PhysRevLett.65.1591}
}
@article{lebo,
  title = {Finite-size effects and shock fluctuations in the asymmetric simple-exclusion process},
  author = {Janowsky, S. A. and Lebowitz, J. L.},
  journal = {Phys. Rev. A},
  volume = {45},
  issue = {2},
  pages = {618--625},
  numpages = {0},
  year = {1992},
  month = {Jan},
  publisher = {American Physical Society},
  doi = {10.1103/PhysRevA.45.618},
  url = {https://link.aps.org/doi/10.1103/PhysRevA.45.618}
}
@article{hinsch,
  title = {Bulk-Driven Nonequilibrium Phase Transitions in a Mesoscopic Ring},
  author = {Hinsch, H. and Frey, E.},
  journal = {Phys. Rev. Lett.},
  volume = {97},
  issue = {9},
  pages = {095701},
  numpages = {4},
  year = {2006},
  month = {Aug},
  publisher = {American Physical Society},
  doi = {10.1103/PhysRevLett.97.095701},
  url = {https://link.aps.org/doi/10.1103/PhysRevLett.97.095701}
}
@article{niladri1,
  title = {Nonequilibrium steady states in asymmetric exclusion processes on a ring with bottlenecks},
  author = {Sarkar, N. and Basu, A},
  journal = {Phys. Rev. E},
  volume = {90},
  issue = {2},
  pages = {022109},
  numpages = {8},
  year = {2014},
  month = {Aug},
  publisher = {American Physical Society},
  doi = {10.1103/PhysRevE.90.022109},
  url = {https://link.aps.org/doi/10.1103/PhysRevE.90.022109}
}
@article{tirtha-niladri,
	doi = {10.1088/1742-5468/2015/01/p01024},
	url = {https://doi.org/10.1088/1742-5468/2015/01/p01024},
	year = 2015,
	month = {jan},
	publisher = {{IOP} Publishing},
	volume = {2015},
	number = {1},
	pages = {P01024},
	author = {T. Banerjee and N. Sarkar and A. Basu},
	title = {Generic nonequilibrium steady states in an exclusion process on an inhomogeneous ring},
	journal = {J. Stat. Mech.: Theory Exp}
	}
	@article{tirtha-qkpz,
  title = {Smooth or shock: Universality in closed inhomogeneous driven single file motions},
  author = {Banerjee, Tirthankar and Basu, Abhik},
  journal = {Phys. Rev. Research},
  volume = {2},
  issue = {1},
  pages = {013025},
  numpages = {10},
  year = {2020},
  month = {Jan},
  publisher = {American Physical Society},
  doi = {10.1103/PhysRevResearch.2.013025},
  url = {https://link.aps.org/doi/10.1103/PhysRevResearch.2.013025}
}
@article{reser1,
	doi = {10.1088/1742-5468/2008/06/p06009},
	url = {https://doi.org/10.1088/1742-5468/2008/06/p06009},
	year = 2008,
	month = {jun},
	publisher = {{IOP} Publishing},
	volume = {2008},
	number = {06},
	pages = {P06009},
	author = {D A Adams and B Schmittmann and R K P Zia},
	title = {Far-from-equilibrium transport with constrained resources},
	journal = {J. Stat. Mech.: Theory Exp}
	}
@article{reser2,
	doi = {10.1088/1742-5468/2009/02/p02012},
	url = {https://doi.org/10.1088/1742-5468/2009/02/p02012},
	year = 2009,
	month = {feb},
	publisher = {{IOP} Publishing},
	volume = {2009},
	number = {02},
	pages = {P02012},
	author = {L Jonathan Cook and R K P Zia},
	title = {Feedback and fluctuations in a totally asymmetric simple exclusion process with finite resources},
	journal = {J. Stat. Mech.: Theory Exp}
	}
@article{reser3,
  title = {Competition between multiple totally asymmetric simple exclusion processes for a finite pool of resources},
  author = {Cook, L. Jonathan and Zia, R. K. P. and Schmittmann, B.},
  journal = {Phys. Rev. E},
  volume = {80},
  issue = {3},
  pages = {031142},
  numpages = {12},
  year = {2009},
  month = {Sep},
  publisher = {American Physical Society},
  doi = {10.1103/PhysRevE.80.031142},
  url = {https://link.aps.org/doi/10.1103/PhysRevE.80.031142}
}
@article{traffic1,
  title = {Macroscopic car condensation in a parking garage},
  author = {Ha, M. and den Nijs, M.},
  journal = {Phys. Rev. E},
  volume = {66},
  issue = {3},
  pages = {036118},
  numpages = {11},
  year = {2002},
  month = {Sep},
  publisher = {American Physical Society},
  doi = {10.1103/PhysRevE.66.036118},
  url = {https://link.aps.org/doi/10.1103/PhysRevE.66.036118}
}
@article{brackley,
  title={Multiple phase transitions in a system of exclusion processes with limited reservoirs of particles and fuel carriers},
  author={Brackley, C. A and Ciandrini, L. and Romano, M C.},
  journal={J. Stat. Mech.: Theory Exp},
  volume={2012},
  number={03},
  pages={P03002},
  year={2012},
  publisher={IOP Publishing},
  	doi = {10.1088/1742-5468/2012/03/p03002},
	url = {https://doi.org/10.1088/1742-5468/2012/03/p03002}
}

@article{limited,
  title = {Limited Resources in a Driven Diffusion Process},
  author = {Brackley, C. A. and Romano, M. C. and Grebogi, C. and Thiel, M.},
  journal = {Phys. Rev. Lett.},
  volume = {105},
  issue = {7},
  pages = {078102},
  numpages = {4},
  year = {2010},
  month = {Aug},
  publisher = {American Physical Society},
  doi = {10.1103/PhysRevLett.105.078102},
  url = {https://link.aps.org/doi/10.1103/PhysRevLett.105.078102}
}
@article{lim-driv,
  title = {Slow sites in an exclusion process with limited resources},
  author = {Brackley, C. A. and Romano, M. C. and Thiel, M.},
  journal = {Phys. Rev. E},
  volume = {82},
  issue = {5},
  pages = {051920},
  numpages = {13},
  year = {2010},
  month = {Nov},
  publisher = {American Physical Society},
  doi = {10.1103/PhysRevE.82.051920},
  url = {https://link.aps.org/doi/10.1103/PhysRevE.82.051920}
}
@article{lim-bio1,
  title={The dynamics of supply and demand in mRNA translation},
  author={Brackley, C. A and Romano, M. C. and Thiel, M.},
  journal={PLoS computational biology},
  volume={7},
  number={10},
  pages={e1002203},
  year={2011},
  publisher={Public Library of Science San Francisco, USA}
}
@article{lim-bio2,
  title={Motor protein traffic regulation by supply--demand balance of resources},
  author={Ciandrini, L. and Neri, I. and Walter, J. C. and Dauloudet, O. and Parmeggiani, A.},
  journal={Phys. biol.},
  volume={11},
  number={5},
  pages={056006},
  year={2014},
  publisher={IOP Publishing},
  	doi = {10.1088/1478-3975/11/5/056006},
	url = {https://doi.org/10.1088/1478-3975/11/5/056006}
}

@article{tirtha-lk1,
  title = {Phase coexistence and particle nonconservation in a closed asymmetric exclusion process with inhomogeneities},
  author = {Banerjee, T. and Chandra, A. K. and Basu, A.},
  journal = {Phys. Rev. E},
  volume = {92},
  issue = {2},
  pages = {022121},
  numpages = {9},
  year = {2015},
  month = {Aug},
  publisher = {American Physical Society},
  doi = {10.1103/PhysRevE.92.022121},
  url = {https://link.aps.org/doi/10.1103/PhysRevE.92.022121}
}
@article{tirtha-lk2,
  title = {Nonequilibrium steady states in a closed inhomogeneous asymmetric exclusion process with generic particle nonconservation},
  author = {Daga, B. and Mondal, S. and Chandra, A.  K. and Banerjee, T. and Basu, A.},
  journal = {Phys. Rev. E},
  volume = {95},
  issue = {1},
  pages = {012113},
  numpages = {9},
  year = {2017},
  month = {Jan},
  publisher = {American Physical Society},
  doi = {10.1103/PhysRevE.95.012113},
  url = {https://link.aps.org/doi/10.1103/PhysRevE.95.012113}
}

@article{astik-1tasep,
  title = {Asymmetric exclusion processes with fixed resources: Reservoir crowding and steady states},
  author = {Haldar, A. and Roy, P. and Basu, A.},
  journal = {Phys. Rev. E},
  volume = {104},
  issue = {3},
  pages = {034106},
  numpages = {12},
  year = {2021},
  month = {Sep},
  publisher = {American Physical Society},
  doi = {10.1103/PhysRevE.104.034106},
  url = {https://link.aps.org/doi/10.1103/PhysRevE.104.034106}
}
@article{blythe,
	doi = {10.1088/1751-8113/40/46/r01},
	url = {https://doi.org/10.1088/1751-8113/40/46/r01},
	year = 2007,
	month = {oct},
	publisher = {{IOP} Publishing},
	volume = {40},
	number = {46},
	pages = {R333--R441},
	author = {R A Blythe and M R Evans}, 
	title = {Nonequilibrium steady states of matrix-product form: a solver{\textquotesingle}s guide},
	journal = {J. Phys. A}
	}


@article{tasep-mft,
title={ refernce on TASEP MFT}
}
@article{lim-exp1,
  title={Cytoplasmic volume modulates spindle size during embryogenesis},
  author={Good, M. C. and Vahey, M. D. and Skandarajah, A. and Fletcher, D. A. and Heald, R.},
  journal={Science},
  volume={342},
  number={6160},
  pages={856--860},
  year={2013},
  publisher={American Association for the Advancement of Science}
}
@article{lim-exp2,
  title={Changes in cytoplasmic volume are sufficient to drive spindle scaling},
  author={Hazel, J. and Krutkramelis, K. and Mooney, P. and Tomschik, M. and Gerow, K. and Oakey, J. and Gatlin, JC},
  journal={Science},
  volume={342},
  number={6160},
  pages={853--856},
  year={2013},
  publisher={American Association for the Advancement of Science}
}
@article{tasep-rev1,
  title={Exactly solvable models for many-body systems far from equilibrium},
  author={Schutz, GM},
  journal={Phase transitions and critical phenomena},
  year={2000}
}
@article{tasep-rev2,
title = {An exactly soluble non-equilibrium system: The asymmetric simple exclusion process},
journal = {Physics Reports},
volume = {301},
number = {1},
pages = {65-83},
year = {1998},
issn = {0370-1573},
doi = {https://doi.org/10.1016/S0370-1573(98)00006-4},
url = {https://www.sciencedirect.com/science/article/pii/S0370157398000064},
author = {B. Derrida}
}
@article{motor1,
  title = {Fluctuation-Driven Molecular Transport Through an Asymmetric Membrane Channel},
  author = {Kosztin, I. and Schulten, K.},
  journal = {Phys. Rev. Lett.},
  volume = {93},
  issue = {23},
  pages = {238102},
  numpages = {4},
  year = {2004},
  month = {Nov},
  publisher = {American Physical Society},
  doi = {10.1103/PhysRevLett.93.238102},
  url = {https://link.aps.org/doi/10.1103/PhysRevLett.93.238102}
}
@article{erwin-lk-prl,
  title = {Phase Coexistence in Driven One-Dimensional Transport},
  author = {Parmeggiani, A. and Franosch, T. and Frey, E.},
  journal = {Phys. Rev. Lett.},
  volume = {90},
  issue = {8},
  pages = {086601},
  numpages = {4},
  year = {2003},
  month = {Feb},
  publisher = {American Physical Society},
  doi = {10.1103/PhysRevLett.90.086601},
  url = {https://link.aps.org/doi/10.1103/PhysRevLett.90.086601}
}
@article{erwin-lk-pre,
  title = {Totally asymmetric simple exclusion process with Langmuir kinetics},
  author = {Parmeggiani, A. and Franosch, T. and Frey, E.},
  journal = {Phys. Rev. E},
  volume = {70},
  issue = {4},
  pages = {046101},
  numpages = {20},
  year = {2004},
  month = {Oct},
  publisher = {American Physical Society},
  doi = {10.1103/PhysRevE.70.046101},
  url = {https://link.aps.org/doi/10.1103/PhysRevE.70.046101}
}
@article{erwin-bottleneck,
  title = {Bottleneck-induced transitions in a minimal model for intracellular transport},
  author = {Pierobon, P. and Mobilia, M. and Kouyos, R. and Frey, E.},
  journal = {Phys. Rev. E},
  volume = {74},
  issue = {3},
  pages = {031906},
  numpages = {13},
  year = {2006},
  month = {Sep},
  publisher = {American Physical Society},
  doi = {10.1103/PhysRevE.74.031906},
  url = {https://link.aps.org/doi/10.1103/PhysRevE.74.031906}
}
@article{tobias-ef1,
  title = {Exclusion Processes with Internal States},
  author = {Reichenbach, T. and Franosch, T. and Frey, E.},
  journal = {Phys. Rev. Lett.},
  volume = {97},
  issue = {5},
  pages = {050603},
  numpages = {4},
  year = {2006},
  month = {Aug},
  publisher = {American Physical Society},
  doi = {10.1103/PhysRevLett.97.050603},
  url = {https://link.aps.org/doi/10.1103/PhysRevLett.97.050603}
}
@article{tobias-ef2,
doi = {10.1088/1367-2630/9/6/159},
url = {https://dx.doi.org/10.1088/1367-2630/9/6/159},
year = {2007},
month = {jun},
publisher = {},
volume = {9},
number = {6},
pages = {159},
author = {T. Reichenbach and E. Frey and T. Franosch},
title = {Traffic jams induced by rare switching events in two-lane transport},
journal = {New Journal of Physics}
}
@article{tobias-ef3,
  title={Domain wall delocalization, dynamics and fluctuations in an exclusion process with two internal states},
  author={Reichenbach, T. and Franosch, T. and Frey, E.},
  journal={Eur. Phys. J. E},
  volume={27},
  pages={47--56},
  year={2008},
  publisher={Springer}
}
@article{klumpp2,
 author ={ Klumpp, S. and Lipowsky, R.},
 year ={ 2003},
 title ={ Traffic of Molecular Motors Through Tube-Like Compartments},
 journal = {J Stat Phys},
 pages = { 233-- 268},
 vol = {113},
 number ={ 1},
 url= { https://doi.org/10.1023/A:1025778922620},
 doi = { 10.1023/A:1025778922620}
 }
 @article{klumpp3,
  title = {Asymmetric simple exclusion processes with diffusive bottlenecks},
  author = {Klumpp, Stefan and Lipowsky, Reinhard},
  journal = {Phys. Rev. E},
  volume = {70},
  issue = {6},
  pages = {066104},
  numpages = {9},
  year = {2004},
  month = {Dec},
  publisher = {American Physical Society},
  doi = {10.1103/PhysRevE.70.066104},
  url = {https://link.aps.org/doi/10.1103/PhysRevE.70.066104}
}
 @article{graf1,
  title = {Generic Transport Mechanisms for Molecular Traffic in Cellular Protrusions},
  author = {Graf, I. R. and Frey, E.},
  journal = {Phys. Rev. Lett.},
  volume = {118},
  issue = {12},
  pages = {128101},
  numpages = {6},
  year = {2017},
  month = {Mar},
  publisher = {American Physical Society},
  doi = {10.1103/PhysRevLett.118.128101},
  url = {https://link.aps.org/doi/10.1103/PhysRevLett.118.128101}
}
@article{graf2,
  title = {Self-organized system-size oscillation of a stochastic lattice-gas model},
  author = {Bojer, M. and Graf, I. R. and Frey, E.},
  journal = {Phys. Rev. E},
  volume = {98},
  issue = {1},
  pages = {012410},
  numpages = {19},
  year = {2018},
  month = {Jul},
  publisher = {American Physical Society},
  doi = {10.1103/PhysRevE.98.012410},
  url = {https://link.aps.org/doi/10.1103/PhysRevE.98.012410}
}


@article{klumpp1,
  title = {Random Walks of Cytoskeletal Motors in Open and Closed Compartments},
  author = {Lipowsky, R. and Klumpp, S. and Nieuwenhuizen, T. M.},
  journal = {Phys. Rev. Lett.},
  volume = {87},
  issue = {10},
  pages = {108101},
  numpages = {4},
  year = {2001},
  month = {Aug},
  publisher = {American Physical Society},
  doi = {10.1103/PhysRevLett.87.108101},
  url = {https://link.aps.org/doi/10.1103/PhysRevLett.87.108101}
}

@article{howard,
    title={Mechanics of motor proteins and the cytoskeleton},
  author={Howard, J. and Clark, RL},
  journal={Appl. Mech. Rev.},
  volume={55},
  number={2},
  pages={B39--B39},
  year={2002}
}
@article{melbinger,
  title={Driven Transport on Parallel Lanes with Particle Exclusion and Obstruction},
  author={A. Melbinger, T. Reichenbach, T. Franosch, and E. Frey},
  journal={Phys. Rev. E},
  volume={83},
  pages={031923},
  year={2011}
 } 
 @article{minm1,
  title = {Minimal current phase and universal boundary layers in driven diffusive systems},
  author = {Hager, J. S. and Krug, J. and Popkov, V. and Sch\"utz, G. M.},
  journal = {Phys. Rev. E},
  volume = {63},
  issue = {5},
  pages = {056110},
  numpages = {12},
  year = {2001},
  month = {Apr},
  publisher = {American Physical Society},
  doi = {10.1103/PhysRevE.63.056110},
  url = {https://link.aps.org/doi/10.1103/PhysRevE.63.056110}
}
@article{rand1,
title= {The Asymmetric Exclusion Process: Comparison of Update Procedures},
author={ Rajewsky, N. and  Santen, L. and  Schadschneider, A. and  Schreckenberg, M.},
year= {1998},
journal={Journal of Statistical Physics},
pages= {151-194},
volume={92},
issue= {1},
URL= {https://doi.org/10.1023/A:1023047703307},
doi={ 10.1023/A:1023047703307}
}

@article{andrea1,
  title = {Understanding totally asymmetric simple-exclusion-process transport on networks: Generic analysis via effective rates and explicit vertices},
  author = {Embley, Ben and Parmeggiani, Andrea and Kern, Norbert},
  journal = {Phys. Rev. E},
  volume = {80},
  issue = {4},
  pages = {041128},
  numpages = {22},
  year = {2009},
  month = {Oct},
  publisher = {American Physical Society},
  doi = {10.1103/PhysRevE.80.041128},
  url = {https://link.aps.org/doi/10.1103/PhysRevE.80.041128}
}

@article{rakesh1,
  title = {Phase transition and phase coexistence in coupled rings with driven exclusion processes},
  author = {Chatterjee, R. and Chandra, A. K. and Basu, A.},
  journal = {Phys. Rev. E},
  volume = {87},
  issue = {3},
  pages = {032157},
  numpages = {7},
  year = {2013},
  month = {Mar},
  publisher = {American Physical Society},
  doi = {10.1103/PhysRevE.87.032157},
  url = {https://link.aps.org/doi/10.1103/PhysRevE.87.032157}
}
@article{rakesh2,
  title={Asymmetric exclusion processes on a closed network with bottlenecks},
  author={Chatterjee, R. and Chandra, A. K. and Basu, A.},
  journal={J. Stat. Mech.: Theory and Exp.},
  volume={2015},
  number={1},
  pages={P01012},
  year={2015},
  publisher={IOP Publishing}
}
@article{anjan-parna,
  title = {Pinned or moving: States of a single shock in a ring},
  author = {Roy, P. and Chandra, A. K. and Basu, A.},
  journal = {Phys. Rev. E},
  volume = {102},
  issue = {1},
  pages = {012105},
  numpages = {16},
  year = {2020},
  month = {Jul},
  publisher = {American Physical Society},
  doi = {10.1103/PhysRevE.102.012105},
  url = {https://link.aps.org/doi/10.1103/PhysRevE.102.012105}
}

@article{atri2,
  title = {Nonequilibrium steady states in coupled asymmetric and symmetric exclusion processes},
  author = {Goswami, A. and Dey, U. and Mukherjee, S.},
  journal = {Phys. Rev. E},
  volume = {108},
  issue = {5},
  pages = {054122},
  numpages = {18},
  year = {2023},
  month = {Nov},
  publisher = {American Physical Society},
  doi = {10.1103/PhysRevE.108.054122},
  url = {https://link.aps.org/doi/10.1103/PhysRevE.108.054122}
}
@article{arvind1,
  title = {Reservoir crowding in a resource-constrained exclusion process with a dynamic defect},
  author = {Pal, B. and Gupta, A. K.},
  journal = {Phys. Rev. E},
  volume = {106},
  issue = {4},
  pages = {044130},
  numpages = {13},
  year = {2022},
  month = {Oct},
  publisher = {American Physical Society},
  doi = {10.1103/PhysRevE.106.044130},
  url = {https://link.aps.org/doi/10.1103/PhysRevE.106.044130}
}

@article{arvind2,
  title = {Exclusion process with scaled resources: Delocalized shocks and interplay of reservoirs},
  author = {Pal, B. and Gupta, A. K.},
  journal = {Phys. Rev. E},
  volume = {105},
  issue = {5},
  pages = {054103},
  numpages = {10},
  year = {2022},
  month = {May},
  publisher = {American Physical Society},
  doi = {10.1103/PhysRevE.105.054103},
  url = {https://link.aps.org/doi/10.1103/PhysRevE.105.054103}
}
@article{arvind3,
  title = {Interplay of reservoirs in a bidirectional system},
  author = {Gupta, A. and Pal, B. and Gupta, A. K.},
  journal = {Phys. Rev. E},
  volume = {107},
  issue = {3},
  pages = {034103},
  numpages = {19},
  year = {2023},
  month = {Mar},
  publisher = {American Physical Society},
  doi = {10.1103/PhysRevE.107.034103},
  url = {https://link.aps.org/doi/10.1103/PhysRevE.107.034103}
}
@misc{sourav1,
      title={Availability, storage capacity, and diffusion: Stationary states of an asymmetric exclusion process connected to two reservoirs}, 
      author={S. Pal and P. Roy and A. Basu},
      year={2024},
      eprint={2308.08384},
      archivePrefix={arXiv},
      url={https://arxiv.org/abs/2308.08384}, 
}
@misc{sourav2,
      title={Distributed fixed resources exchanging particles: Phases of an asymmetric exclusion process connected to two reservoirs}, 
      author={S. Pal and P. Roy and A. Basu},
      year={2024},
      eprint={2403.05945},
      archivePrefix={arXiv},
      url={https://arxiv.org/abs/2403.05945}, 
}
@article{ciandrini,
  title = {Driven transport on a flexible polymer with particle recycling: A model inspired by transcription and translation},
  author = {Fernandes, Lucas D. and Ciandrini, Luca},
  journal = {Phys. Rev. E},
  volume = {99},
  issue = {5},
  pages = {052409},
  numpages = {10},
  year = {2019},
  month = {May},
  publisher = {American Physical Society},
  doi = {10.1103/PhysRevE.99.052409},
  url = {https://link.aps.org/doi/10.1103/PhysRevE.99.052409}
}
@article{dauloudet,
  title={Modelling the effect of ribosome mobility on the rate of protein synthesis},
  author={Dauloudet, O. and Neri, I. and Walter, J-C and Dorignac, J. and Geniet, F. and Parmeggiani, A.},
  journal={ Eur. Phys. J. E},
  volume={44},
  pages={1-15},
  year={2021},
  publisher={Springer}
}
@article{atri1,
      title={Defect versus defect: stationary states of single file marching in periodic landscapes with road blocks}, 
      author={A. Goswami and R. Chatterjee and S. Mukherjee},
      year={2024},
      eprint={2402.08499},
      archivePrefix={arXiv},
      primaryClass={cond-mat.stat-mech},
      url={https://arxiv.org/abs/2402.08499}, 
}
@article{melbing1,
  title = {Microtubule Length Regulation by Molecular Motors},
  author = {Melbinger, A. and Reese, L. and Frey, E.},
  journal = {Phys. Rev. Lett.},
  volume = {108},
  issue = {25},
  pages = {258104},
  numpages = {5},
  year = {2012},
  month = {Jun},
  publisher = {American Physical Society},
  doi = {10.1103/PhysRevLett.108.258104},
  url = {https://link.aps.org/doi/10.1103/PhysRevLett.108.258104}
}

@article{howard1,
doi:10.1073/pnas.1107281109,
author = {Cécile Leduc  and Kathrin Padberg-Gehle  and Vladimír Varga  and Dirk Helbing  and Stefan Diez  and Jonathon Howard },
title = {Molecular crowding creates traffic jams of kinesin motors on microtubules},
journal = { Proc. Natl. Acad. Sci. },
volume = {109},
number = {16},
pages = {6100-6105},
year = {2012},
doi = {10.1073/pnas.1107281109},
URL = {https://www.pnas.org/doi/abs/10.1073/pnas.1107281109}
}
@article{melbing2,
  title={Crowding of molecular motors determines microtubule depolymerization},
  author={Reese, Louis and Melbinger, Anna and Frey, Erwin},
  journal={Biophysical journal},
  volume={101},
  number={9},
  pages={2190--2200},
  year={2011},
  publisher={Elsevier}
}

\maketitle

\section{Introduction}

Driven diffusive systems~\cite{driven-diff1,driven-diff2,driven-diff3,driven-diff4}   have served as simple models where
general features of nonequilibrium steady states and phase transitions have been studied.
Already in one spatial dimension, such driven systems show highly non-trivial properties even with purely local dynamics. 
A paradigmatic example is the totally asymmetric simple exclusion
process (TASEP)~\cite{krug,krug1,krug2,derrida}. 
It consists of a  one-dimensional (1D) lattice with open boundaries. 
The dynamics are subject to exclusion at all sites, i.e., they can accommodate a maximum of one particle per site and is stochastic in nature. It involves entering a particle at one of the boundaries at a specified rate, unidirectional hopping to the following sites at rate unity until it reaches the other end, from which it leaves at another specified rate. 
As a function of the entry and exit rates at the boundary, the TASEP exhibits three distinct phases: 
the steady-state densities in bulk can be either less than half, giving the low-density phase, or more than half, giving the high-density phase, or just half, which is the maximal current phase. 


Historically, the TASEP was proposed as a simple model to describe protein synthesis in biological cells~\cite{macdonald}.
It became a paradigmatic nonequilibrium model when it was discovered to have boundary-induced phase transitions~\cite{krug,krug1,krug2}. 
{Motivated by various phenomena across biological and social systems, TASEP has been extended and generalized, leading to the discovery of novel phases and collective phenomena.}
For instance, it have been discovered that accounting for the exchange of particles between the transport lane and the environment gives rise to traffic jams~\cite{erwin-lk-prl,erwin-lk-pre}. 
The theoretical prediction of traffic jams has recently been nicely confirmed experimentally~\cite{howard1}.
These studies have been extended further and it was shown how molecular motors from the kinesin-8 family can regulate microtubule depolymerization dynamics and ultimately lead to length control of microtubules~\cite{melbing1,melbing2}.
Additionally, intriguing behaviors such as domain wall delocalization and traffic jams are found in systems with two TASEP lanes coupled through rare particle-switching events~\cite{tobias-ef1,tobias-ef2,tobias-ef3}.

Inspired by the dynamics of molecular motors along complex networks of microtubules in biological cells~\cite{howard}, TASEP has been explored on simple networks, studying the conditions for the existence of the various phases and domain walls~\cite{andrea1,rakesh1,rakesh2} on various segments of the networks. 
Another area of research on TASEP focuses on how the interplay between TASEP on an inhomogeneous ring and particle number conservation can generate macroscopically nonuniform steady states. Studies in this direction have explored phenomena such as domain walls, their fluctuation-induced delocalization, and the competition between different types of defects~\cite{lebo,hinsch,niladri1,tirtha-niladri,tirtha-lk1,tirtha-lk2,tirtha-qkpz,anjan-parna,atri1}.

Another interesting variant of TASEP involves coupling it with the symmetric exclusion process (SEP).
This coupling has been studied in both half-closed~\cite{graf1,graf2} and fully open geometries~\cite{atri2}. Some of the notable results from these studies are tip  localization of motors~\cite{graf1}, self-organized temporal patterns~\cite{graf2} and  complex space-dependent stationary states in the both TASEP and SEP lanes~\cite{atri2}.

 To model the coupled system of molecular motors and microtubules,  in approaches complementary to Refs.~\cite{erwin-lk-prl,erwin-lk-pre}, a TASEP lane has been considered to be confined in and coupled to an embedding three-dimensional (3D) particle reservoir via particle exchanges. 
The consequences of the effects of diffusion on
the steady states of the filament (TASEP) have been explored in Refs.
~\cite{klumpp1,klumpp2,ciandrini,dauloudet}, 
giving physical insights on the nontrivial steady states arising out of coupling between a 1D driven (TASEP) and 3D equilibrium (diffusion) processes. Further, related to these studies but staying within a strict 1D description, Ref.~\cite{klumpp3} considered a 1D open model with diffusive and driven segments connected serially, and explored the resulting modifications of the pure open TASEP phase diagram.   

In the present study, {we examine a distinct class of TASEP models that differ significantly from those previously discussed. These models do not feature open boundaries but instead connect the TASEP to a particle reservoir at both ends, imposing a constraint on the total number of particles in the combined system. This 'restricted access' to a finite number of particles effectively limits the particle population within the system, distinguishing it from other TASEP models.}
{This class of models is inspired by biological processes such as protein synthesis, where the finite availability of ribosomes in a cell limits the number of proteins synthesized~\cite{reser1}. 
Similarly, it reflects traffic dynamics, where the finite number of vehicles affects the flow within a network of roads~\cite{traffic1}; see also Ref.~\cite{limited} 
in this context.}
{Examples of TASEP models connected to a reservoir include  driven diffusive systems with limited resources~\cite{lim-driv}, various biological systems such as mRNA translation and motor protein dynamics in cells~\cite{lim-bio1}, and traffic flow problems~\cite{lim-bio2}.}

In the studies on TASEPs with finite resources, the actual particle entry rates, unlike open TASEPs, depend upon the instantaneous reservoir population; the exit rates are assumed to be constants, similar to open TASEPs. 
A closed system consisting of a TASEP and a reservoir  manifestly breaks the translation invariance and hence is a potential candidate for inhomogeneous steady-state densities. 
How the finite pool of available particles affect the steady state occupations and currents in comparison with open TASEPs is the key qualitative question in all these studies. 

At the simplest level of modeling of a TASEP with finite resources, a single TASEP is connected at both ends to a particle reservoir, with the entry rate to the TASEP lane from the reservoir depending on the instantaneous reservoir occupation; 
for simplicity, the reservoir is generally taken to be a point reservoir without any spatial extent.
More than one TASEP connected to a reservoir has also been considered that allows us to study competition between TASEPs for finite resources; see Ref.~\cite{reser3}. 
Detailed studies, both numerical stochastic simulations and analytical mean-field theory, reveal rich nonuniform steady-state density profiles, including domain walls in these 
models~\cite{reser1,reser2,reser3}. 
In yet another extension of this problem, the effect of a limited number of two different fundamental transport resources, \textit{viz.} the hopping particles and the ``fuel carriers'', which supply the energy required to drive the system away from equilibrium has been studied~\cite{brackley}, and multiple phase transitions are reported as the entry and exit rates are changed.  
It may be noted that in all these studies on TASEPs with finite resources, the effective entry rate of the particles from the reservoir to the TASEP lane is not constant but depends on the instantaneous reservoir population. 
In contrast, the exit rate from the TASEP to the reservoir is taken to be constant, as in an open TASEP. 
Very recently, in substantially different modeling from the above, both the entry and exit rates are assumed to depend on the instantaneous reservoir population leading to dynamics-induced competition between the entry and exit rates~\cite{astik-1tasep}. 
How that determines the steady states of a TASEP connected to a finite reservoir has been studied, revealing significant differences in the phase diagram of the model with the other existing model of a TASEP with finite resources~\cite{astik-1tasep}, which highlights the role of the competition between the effective entry and exit rates. { {This study has been extended by incorporating a dynamic defect in the TASEP lane~\cite{arvind1}. More recently, some studies have considered  TASEPs with
 multiple reservoirs. These include studies on the interplay of
 two reservoirs coupled to two TASEPs~\cite{arvind2} and interplay of
 reservoirs in a bidirectional system~\cite{arvind3}. Very recently, how direct particle exchanges between two reservoirs connected to a TASEP affect the stationary densities on the TASEP lane have been considered in Refs.~\cite{sourav1,sourav2}, giving insights on the competition between particle exchanges and TASEP. }}


Despite the recent impressive progress in this research, many open questions are yet to be addressed or investigated systematically.  { {In the above-mentioned studies on TASEPs connected to finite reservoirs, the reservoirs are assumed to be point-like, without any spatial extent. However,
 real systems can be spatially extended, where different regions may act as carrying resources for the whole system. }}
This evidently should lead to currents between different regions in the system. 
Furthermore, these currents may not only be controlled by the availability of resources at the ``supply side'', but also by the carrying capacity of the ``receiving side''. 
In fact, the interplay between the finite availability and carrying capacity of particles at different parts of the system, together with the global conservation of the particles, should be the key determining factors of the eventual nonequilibrium steady state in such a system. 
Clearly, a single TASEP attached to a single reservoir cannot capture these phenomenologies. { {Some research has been done broadly along this line recently. These include studies on the interplay of
 reservoirs in a bidirectional system~\cite{arvind3}. Very recently, how direct particle exchanges between two reservoirs connected to a single TASEP lane affect the stationary densities on the TASEP lane have been considered in Refs.~\cite{sourav1,sourav2}, giving insights on the competition between particle exchanges and TASEP.  }}

The general goal of the present study is to theoretically understand the classes of nonequilibrium steady states (NESS) in a system with multiple reservoirs having finite capacities and overall particle number conservation.  
In particular, being inspired by the physics of the interplay between the finite availability and carrying capacity of particles at different parts of the system in conjunction with the global conservation of the particles, we construct and study a simple conceptual model that consists of two reservoirs having the same finite capacity without any internal dynamics and two TASEP lanes with equal size connecting the two reservoirs in an antiparallel manner (see below) that ensures a finite steady-state current.  
We show that for appropriate choices of model parameters, this model can admit uniform densities in the TASEP lanes having equal or different values, as well as spatially nonuniform densities in the form of static or moving domain walls. 
Further, the population of the two reservoirs can also be controlled in our model - in fact, by tuning the model parameters, the reservoirs can be preferentially populated or depopulated separately in the steady states. This implies that the dynamics can maintain steady macroscopic population imbalances across the system.

\textit{Outline---}
The remainder of this article is organized as follows. Section~\ref{model} introduces the model and defines the model parameters. 
Then in Section~\ref{mean-field theory}, we illustrate the complex phase behavior of the model in a reduced subspace of the entire parameter space using a mean-field description that we develop and supplemented by extensive Monte-Carlo simulations (stochastic simulations). 
Section~\ref{ph-trans} discusses the phase transitions and their order in this model. 
 We summarize our results and discuss their implications in Section~\ref{summ}. In Appendix~\ref{sim-algo}, we briefly discuss our simulation algorithm for our stochastic simulations studies. 
We discuss some additional results on the phase diagrams in Appendix.

\section{Model}\label{model}

\begin{figure}[htb]
\includegraphics[width=\columnwidth]{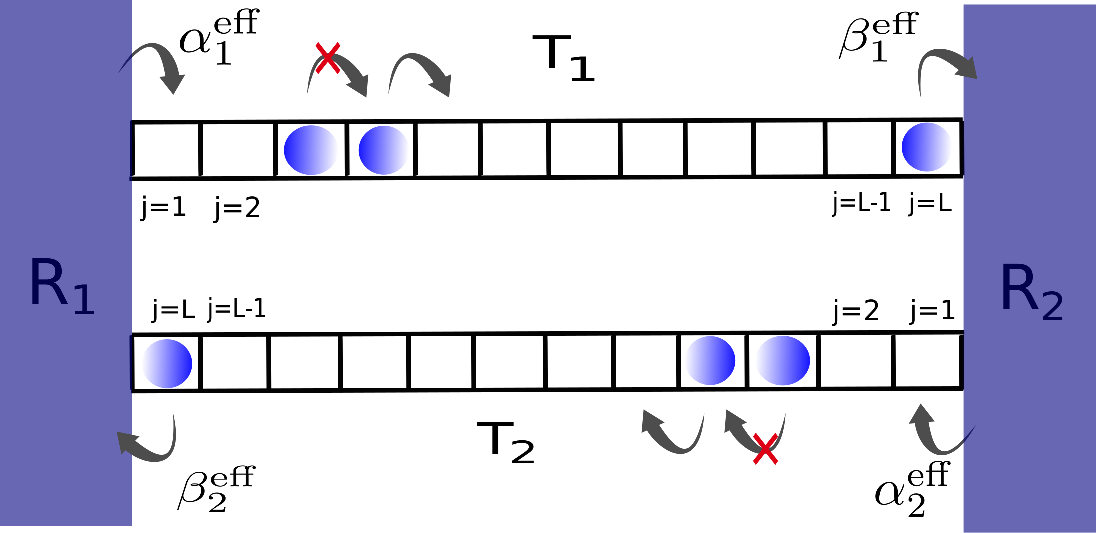}
\caption{
\textbf{Illustration of the model.} 
Two anti-parallel TASEP lanes $T_1$ and $T_2$ of equal length $L$ connect two reservoirs $R_1$ and $R_2$ of equal carrying capacity. 
The effective entrance rates from $R_i$ to $T_i$ at site ${j=1}$ are denoted by $\alpha_i^\text{eff}$, and correspondingly $\beta_i^\text{eff}$ denote the effective exit rate of particles from $T_i$ at site ${j=L}$. 
Particles move unidirectionally along both lanes subject to site exclusion; the hopping rates are assumed to be the same in both lanes and are set to unity to fix the time scale.
}
\label{schem}
\end{figure}

We consider a model with two transport lanes $T_1$ and $T_2$ of equal length $L$ connected to two particle reservoirs $R_1$ and $R_{2}$ [Fig.~\ref{schem}]. 
Transport along the two lanes is anti-parallel, where on each lane, particles move unidirectionally subject to site exclusion, i.e., the particle dynamics is described by a totally asymmetric simple exclusion process (TASEP). 
{For simplicity, we assume that the hopping rates are the same on both lanes and choose them as $1$ to set the time scale.}
We assume that the carrying capacities of the two well-mixed reservoirs are finite and equal, and particle exchange between these reservoirs is exclusively through transport along the two lanes. 
Due to the closed geometry of the system, there is thus an overall limitation of resources. 

Let $N_0$ be the total number of particles available, and $N_{1,2}$ the number of particles in each reservoir $R_{1,2}$. 
Then particle number conservation implies 
\begin{equation}
    N_0
    =
    N_1
    + N_2
    + \sum_{j=1}^L 
      \big(\rho^1_{j} +\rho^2_{j}\big) \, ,
\end{equation}
where ${\rho^1_{j} \in \{0,1\}}$ and ${\rho^2_{j} \in \{0,1\}}$ are the occupation numbers at site $j$ of $T_1$ and $T_2$, respectively. Here and below, superscripts 1 and 2 refer to the two TASEP lanes, and a subscript refers to a lattice site $j$ in a given TASEP lane; see Fig.~\ref{schem}.

The exchanges of particles between the reservoirs and the lanes are characterized by entry and exit rates, which respect both the fact that the available resources are limited and that the carrying capacity of each reservoir is limited. 
To account for resource limitations, we assume for the entry rates from reservoir $R_i$ to lane $T_i$
\begin{align}
    \alpha_{i}^\text{eff} (N_i) 
    = 
    \alpha_i \, f (N_i) 
    \, , 
    \label{def1}
\end{align}
where $\alpha_i$ are positive constants and $f(N_i)$ is a monotonically increasing function of the number of available particle resources $N_i$ in reservoir $R_i$; 
in particular, one requires ${f(0)=0}$. 
For simplicity, we use a linear dependence [see, e.g., Ref.~\cite{sourav1} in this context]
\begin{align}
    f(N_i) = \frac{N_i}{L} \, .\label{def2}
\end{align}
The finite carrying capacity for each reservoir must be considered when defining the exit rates. 
We assume that the  exit rate from lane $T_1$ into reservoir $R_2$ is given by
\begin{align}
    \beta_{1}^\text{eff} = \beta_1 \, g(N_2) \, ,
\label{effrate}
    \end{align}
where $g$ is a monotonically decreasing function of the number of particles $N_2$ already contained in the reservoir $R_2$; similarly, we assume 
\begin{equation}
{\beta_{2}^\text{eff} = \beta_2 \, g(N_1)}.\label{effrate2}
\end{equation}
Again, for simplicity, we take a linear dependence
\begin{align}
    g(N_i) 
    = 
    1 - f(N_i) = 1 - \frac{N_i}{L} 
    \, .
\label{def4}
\end{align}
Here, we specifically have assumed that the carrying capacities of each reservoir are limited to ${N_i \leq L}$. 
Since the total carrying capacity of the two TASEP lanes is $2L$, the maximum number of particles in the system is ${N_0^\text{max} = 4 L}$.
We define the \textit{filling factor}
\begin{align}
    \mu = \frac{N_0}{2L} \, ,
\end{align}
whose maximum value is ${\mu^\text{max}=2}$; 
at ${\mu=1}$ the system is half-filled. 
Particle number conservation then implies
\begin{align}
    N_1+N_2 
    + 
    \sum_{j=1}^L
    \big( 
    \rho^1_{j}+\rho^2_{j}
    \big) 
    = 2\mu L 
    \, .
\label{total-part}
\end{align}
While all these particular choices for the functions $f$ and $g$ characterizing resource limitations and carrying capacities of the reservoirs are obviously dictated by their mathematical simplicity, we expect our conclusion to apply qualitatively to more general cases.

Taken together, our model is characterized by five control parameters, the two entry rates $\alpha_i$, the two exit rates $\beta_i$, and the filling fraction $\mu$. 
For simplicity of analysis, we only consider the symmetric case where ${\alpha_1 = \beta_2 =:a}$ and ${\alpha_2 = \beta_1 =:b}$.
In this case, the dynamics remain invariant under the interchanges ${a\leftrightarrow b}$, ${T_1 \leftrightarrow T_2}$, and ${R_1\leftrightarrow R_2}$ concomitant with ${\rho_1 \leftrightarrow \rho_2}$ and ${N_1\leftrightarrow N_2}$.

\section{Mean-field theory}\label{mft}


Next, we analyze the stochastic model at the mean-field level following the same reasoning as in earlier work; 
for a review see Ref.~\cite{blythe}. 
Neglecting correlations and factorizing higher order correlation functions the equations of motion for the mean densities $\rho^1_{j}(t)$ and $\rho^2_{j}(t)$ read for any site ${j \in [2,L{-}1]}$ in the bulk 
\begin{equation}
    \frac{d}{dt}
    \rho^i_j (t)
    =
    \rho^i_{j-1} 
    \big[ 1 - \rho^i_{j} \big]
    - \rho^i_{j} 
    \big[ 1-\rho^i_{j+1} \big]
    \, .
\label{beq1}
\end{equation}
For sites at the boundaries, we have 
\begin{subequations}
\begin{eqnarray}
    \frac{d}{dt} 
    \rho^i_1(t)
    &=&
    \alpha_i^\text{eff}
    \big[ 1-\rho^i_1 \big] 
    - 
    \rho^i_1 
    \big[ 1-\rho^i_2 \big]
    \, ,
    \label{beq3}
    \\
    \frac{d}{dt} 
    \rho^i_L (t)
    &=&
    \rho^i_{L-1}
    \big[ 1-\rho^i_{L} \big] 
    -
    \beta_i^\text{eff} \rho^i_{L}
    \, .
    \label{beq4}
\end{eqnarray}
\end{subequations}

The model exhibits \textit{particle-hole symmetry} in the following sense: 
In lane $T_1$, a jump of a particle to the \textit{right} corresponds to a vacancy movement by one step to the \textit{left}.
 Similarly, a particle entering $T_1$ at the left boundary can be interpreted as a hole leaving it through the left boundary, and vice versa for the right boundary. 
Likewise, in lane $T_2$, a jump of a particle to the \textit{left} corresponds to a vacancy movement by one step to the \textit{right}.
Furthermore, a particle entering lane $T_2$ at the right boundary may be interpreted as a vacancy leaving it through the right boundary, and vice versa for the left boundary. 
It is easy to check that the transformations
\begin{subequations}
\begin{align}
    \mu 
    &\leftrightarrow 2-\mu
    \, .
    \label{phmu}\\
    \rho^i_j 
    &\leftrightarrow 1-\rho^i_{L-(j-1)}
    \, , 
    \label{ph1} \\
    \alpha_{1,2}^\text{eff} 
    &\leftrightarrow \beta_{1,2}^\text{eff}
    \, , 
    \label{ph3}
\end{align}
\end{subequations}
leave Eq.~\eqref{total-part} and Eqs.~(\ref{beq1})-(\ref{beq4}) invariant. 
Furthermore, since ${\alpha_1^\text{eff} = a N_1/L}$, ${\beta_1^\text{eff} = b (1{-}N_2/L)}$, ${\alpha_2^\text{eff} = b N_2/L}$, and ${\beta_2^\text{eff} = a(1{-}N_1/L)}$, Eq.~\eqref{ph3} implies invariance under ${a\leftrightarrow b}$ together with ${N_1/L \leftrightarrow 1-N_2/L}$. 
These properties allow us to obtain phase diagrams of the model for ${\mu > 1}$ (i.e., more than half-filled) from those with ${\mu < 1}$ (i.e., less than half-filled).

In what follows below, for convenience of presentation of our results, we take taking continuum limit with ${\rho^i(x)\equiv \langle \rho^i_j\rangle}$, denoting the steady state  densities in $T_1$ and $T_2$; 
$x=j/L$ becomes quasi-continuous in the  limit ${L\rightarrow \infty}$ with ${0\leq x\leq 1}$~\cite{hinsch}. For each of $T_1$ and $T_2$, $x$ starts from 0 at the respective entry end with $x=1$ at the respective exit end.

\section{Mean-field phase diagrams}
\label{mean-field theory}



In steady state (within mean-field theory), each of the two lanes is characterized by a constant particle current~\cite{blythe} 
\begin{align}
    J_i = \rho_i \, (1 - \rho_i)
    \, , 
\end{align}
where ${\rho_i \in [0,1]}$ denotes the steady-state particle density on lane $T_i$. 
These currents are related to the number of particles in each reservoir by the flux balance equation
\begin{eqnarray}
    \frac{dN_1}{dt} 
    = J_2-J_1 
    = -\frac{dN_2}{dt}  
    \, .
\label{bas-curr}
\end{eqnarray}
Hence, in steady state, the currents on both lanes must be equal 
\begin{align}
    J = J_1 = J_2 \, .
\end{align}
For a given current $J$, the possible densities are
\begin{equation}
    \rho 
    = 
    \frac{1}{2}
    \left( 1\pm \sqrt{1-4J} \right)
    =: \rho_\pm 
    \, .
\label{rhopm}
\end{equation}
Then, there are the following possible steady states:
\begin{itemize}
    \item \textbf{Uniform and equal densities:} The densities on both lanes are spatially uniform and equal to each other: ${\rho_1 = \rho_2 = \rho}$. For ${\rho = \rho_-}$ and  ${\rho = \rho_+}$, both lanes are in a low density (LD) and a high density (HD) phase, respectively. In the particular case that the current is at its maximal value ${J = \frac14}$, the density equals ${\rho = \frac12}$ (maximal current (MC) phase).
    \item \textbf{Uniform and unequal densities:} The densities on both lanes are spatially uniform but distinct. Then ${\rho_1 = \rho_\pm}$ and ${\rho_2 = 1 - \rho_1 = \rho_\mp}$. This means that while one lane has a bulk density $\rho_+$, the other has $\rho_-$, and vice versa.
    
    \item \textbf{Nonuniform densities:} The density profile on one of the two lanes or both exhibit a domain wall (DW), i.e., a step-like profile connecting the densities $\rho_\pm$ in bulk. These domain walls may be fixed in place (localized) or move (delocalized); see the discussion further below.
\end{itemize}

A list of the distinct phases on lanes $T_1$ and $T_2$ (excluding those obtained by interchanging $T_1$ and $T_2$) are shown in a tabular form in Table~\ref{phase-tab}. 

\begin{table}[h!]
 \begin{center}
  \begin{tabular}{|p{1.5cm}|p{1,5cm}|p{2.5cm}|}
    \hline
    Phases  & $T_1$ & $T_2$ \\
    \hline
    LD-LD & $\rho_1=\rho_-$ & $\rho_2=\rho_-$\\
    HD-HD & $\rho_1=\rho_+$ & $\rho_2=\rho_+$\\
    LD-HD & $\rho_1=\rho_-$ & $\rho_2=\rho_+$ \\
    LD--DW & $\rho_1=\rho_-$ & $\rho_2$: {domain wall} \\
    HD-DW & $\rho_1=\rho_+$ & $\rho_2$: {domain wall} \\
    MC-MC & $\rho_1=1/2 $    & $\rho_2=1/2$\\
    \hline
  \end{tabular}
\caption{List of the distinct phases on lanes $T_1$ and $T_2$. 
This list does not include phases that can be obtained by the interchange of $T_1$ and $T_2$.}
\label{phase-tab}
\end{center}
\end{table}

\subsection{Phase diagrams and phase boundaries}
\label{bound}

\begin{figure*}[htb]
\centering \includegraphics[width=\columnwidth]{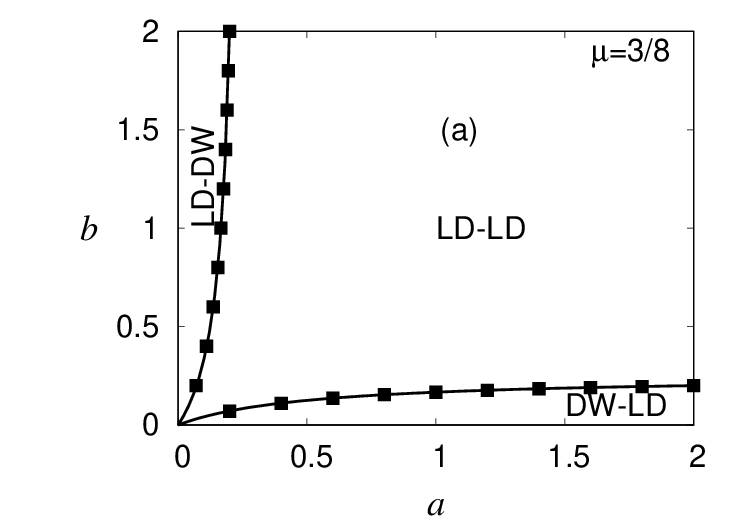} 
 \hfill 
\includegraphics[width=\columnwidth]{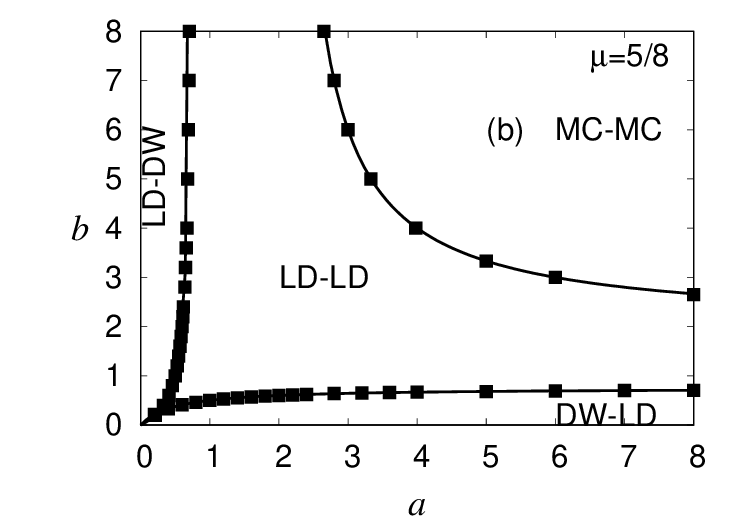}
 \\
\includegraphics[width=\columnwidth]{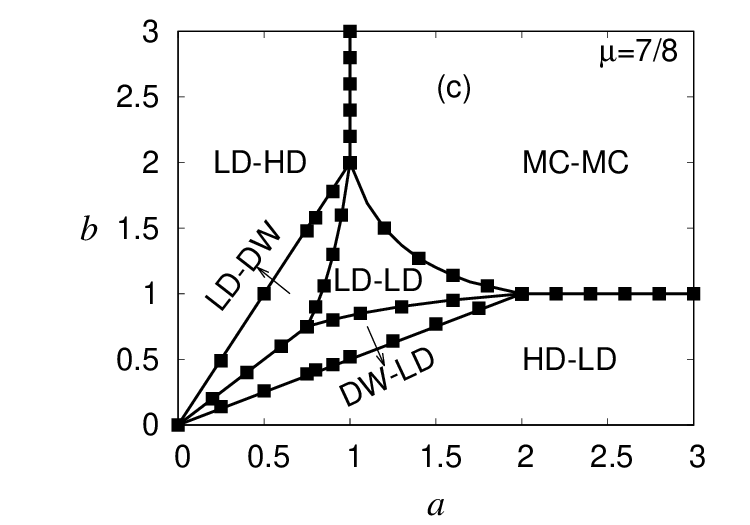}
 \hfill
\includegraphics[width=\columnwidth]{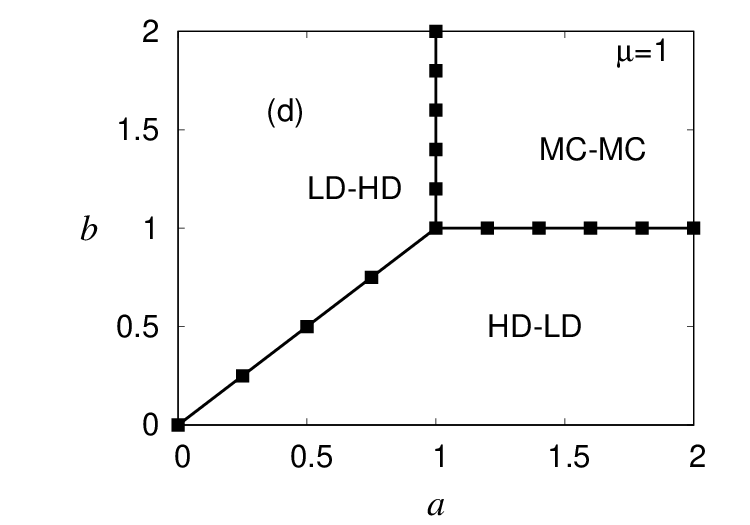}
 \\ 
\caption{
\textbf{Phase diagram} in the $a-b$ plane obtained from mean-field theory (continuous lines) and stochastic simulations (symbols)
for a set of representative filling factors $\mu$ indicated in the graphs. 
(a) $\mu=3/8$: 
The phases are LD--LD, DW--LD, and LD--DW.
This topology of the phase diagram is valid for $1/4< \mu <1/2$; 
(b) $\mu=5/8$: 
The phases are LD-LD, DW-LD, LD--DW and MC-MC, valid for $1/2< \mu <3/4$; 
(c) $\mu=7/8$: 
The phases are LD-LD, LD-HD, HD-LD, DW-LD, LD--DW, and MC-MC, valid for $3/4<\mu <1$; 
(d) $\mu=1$: The phases are LD-HD, HD-LD, and MC-MC. {See the main text for a detailed discussion.}
}
\label{phase_diag}
\end{figure*}


Now that we know the possible phases of the reservoir-coupled TASEP lanes, the next task is to determine where these steady states are located in the space of the control parameters $a$, $b$, and $\mu$.
Figure~\ref{phase_diag} shows the mean-field phase diagram as a function of the two control parameters $a$ and $b$ and a set of representative values for the filling factor $\mu$ ($\mu=3/8,\,5/8,\,7/8,\,1$);
see Fig.~\ref{phase_diag_hole} in Appendix~\ref{phase-great}  for $\mu=9/8,\,11/8,\,13/8$. 
The obtained phase diagrams show different topologies depending on the value of the filling fraction $\mu$,  with a much richer range of phases than an open TASEP. 
Some of these phases and phase diagrams have no analogs in other models for TASEPs  with finite resources~\cite{reser1,reser2} having the same number of control parameters (three).

When determining the phase diagram, we proceeded similar as in Refs.~\cite{reser1,hinsch,melbinger}.
We determine these phases by comparing with the standard phases of an open TASEP, parameterized by $\alpha_T$ and $\beta_T$ as the entry and exit rates, respectively; here, $0<\alpha_T,\,\beta_T<1$. 
Recall that in an open TASEP, the conditions for the transitions between the different phases can be obtained by equating the respective currents in those phases~\cite{blythe}. 
From this condition, one infers that the transition between the LD and HD phases occurs when ${\alpha_T = \beta_T < 1/2}$.
Moreover, since $\alpha_T$ and ${1-\beta_T}$ are the bulk densities in the LD and HD phases, respectively, the above condition also implies that the sum of the bulk densities in the LD and HD phases should be unity at the transition. 
Similarly, one finds that the transitions between the LD and MC phases and HD and MC phases occur, respectively, for ${\alpha_T = 1/2 < \beta_T}$, and ${\beta_T = 1/2 < \alpha_T}$. 
The phase boundary between the LD and HD phases is a first-order line characterized by phase coexistence in the form of a single DW; 
the equation for this line is ${\alpha_T = \beta_T \leq 1/2}$. 
These considerations allow us to obtain the phase diagram of an open TASEP as a function of  $\alpha_T$ and $\beta_T$: For $\alpha_T<1/2,\,\alpha_T<\beta_T$ one has the LD phase, the HD phase exists for $\beta_T<1/2,\,\beta_T<\alpha_T$, and the MC phase is found when $\alpha_T>1/2,\,\beta_T>1/2$.

Since the present model has two TASEP lanes, the phases of both lanes must be specified simultaneously to determine the steady state of the model. 
Again, following the same logic used to obtain the phase diagram for an open TASEP, the transitions between the phases in which $T_1$ and $T_2$ each have a spatially uniform density (which may or may not be equal) can be determined by equating the respective currents.
For instance, this condition tells us that the phase boundary between the LD-HD and the MC-MC phases is given by equating the densities $\rho_1(\text{LD})$ and $\rho_2(\text{HD})$ of $T_1$ and $T_2$ in their LD and HD phases respectively to 1/2: ${\rho_1(\text{LD})=1/2}$ and ${\rho_2(\text{HD})=1/2}$. 
Similarly, the boundary between the LD-LD and MC-MC phases is given by the conditions ${\rho_1(\text{LD})=1/2}$ and ${\rho_2(\text{LD})=1/2}$. 
In Fig.~\ref{phase_diag}, it can further be seen that there are phases with ``domain walls'' (DW) that basically generalize the coexistence line of an open TASEP and are characterized by a single DW with a specific position as a function of $a$, $b$, and $\mu$;
see Ref.~\cite{hinsch} for a similar DW phase in a ring model consisting of a driven and a diffusive segment. 
The boundaries of the DW phases can be determined by considering the positions of the DW  {at the two boundaries of the TASEP lane}~\cite{hinsch}. 
In the following, we will explain the emergence of the different phases and the resulting phase boundaries in more detail. 
 
To validate our mean-field theory, we have also performed stochastic simulations (using a random sequential updating algorithm) to determine the phase diagrams, and we find good agreement with our mean-field theory predictions; 
see Appendix~\ref{sim-algo} for more details on the numerical simulations.
 
\subsection{Uniform lane density: $\rho_1=\rho_2$}
\label{mean-field theory1}

Assuming that the densities on the two lanes are the same and spatially uniform, ${\rho_1 = \rho_2 = \rho}$, the condition of particle number conservation reads 
\begin{equation}
\label{eq:pnc_A}
    N_0
    =
    N_1 + N_2 + 2L\rho 
    \, .
\end{equation}
Clearly, the possible phases are the LD-LD, HD-HD and MC-MC phases. \\

\noindent \textbf{LD-LD phase.}
If both lanes are in their LD phases one has ${\rho_1 = \alpha_1^\text{eff}}$ and ${\rho_2=\alpha_2^\text{eff}}$ \cite{krug}. 
Since ${\rho_1=\rho_2}$, this implies ${a \, f(N_1) = b \, f(N_2)}$, resulting in  
\begin{equation}
    \frac{N_1}{N_2}
    =
    \frac{b}{a}
    \, .
\end{equation}
Together with particle number conservation [Eq.~\eqref{eq:pnc_A}] this yields the number of particles in each of the reservoirs in terms of the filling fraction $\mu$
\begin{equation}
    N_1
    =
    \frac{2L\mu b}{a+b+2ab} 
    \, , 
    \quad
    N_2
    =
    \frac{2L\mu a}{a+b+2ab} 
    \, ,
\label{n1ld}
\end{equation}
and the density on both lanes
\begin{equation}
    \rho_\text{LD}^{} 
    =\frac{2a\mu }{1+2a+a/b} \, .\label{rho-ld-uni}
\end{equation}
As expected, the density of particles in both reservoirs and on both transport lanes scale with the filling fraction $\mu$. \\

\noindent \textbf{HD-HD phase.}
If both lanes are in an HD phase, $\rho_1 = \rho_2 = \rho_\text{HD}^{}$, we have ${\rho_1 = 1-\beta_1^\text{eff}}$ and ${\rho_2 = 1-\beta_2^\text{eff}}$. 
This implies (using the explicit expressions for the effective rates {Eq.~\eqref{effrate}})
\begin{eqnarray}
    a \, \bigg( 1-\frac{N_1}{L} \bigg)
    =
    b \, \bigg( 1-\frac{N_2}{L} \bigg) \, ,
\end{eqnarray}
which, when combined with particle number conservation [Eq.~\eqref{eq:pnc_A}], yields
\begin{subequations}
\begin{align}
    N_1 
    &=\frac{N_0-L(3-2a-a/b)}{1+2a+a/b} 
    \, ,\label{n1hd} \\
    N_2 
    &= {\frac{aN_0/b+L(1+2a-3a/b)}{1+2a+a/b}} 
    \, \label{n2hd}.
\end{align}
\end{subequations}
Thus one obtains for the lane density
\begin{eqnarray}
    \rho_\text{HD}^{}
    =
    \frac{1-2a+a/b+2a\mu }{1+2a+a/b} 
    \, .
\label{rho_hd}
\end{eqnarray}
The above results for the HD-HD phase are clearly consistent with the particle-hole symmetry articulated above, vis-\`a-vis the LD-LD phase. 
\\


\noindent \textbf{MC-MC phase.}
For an MC-MC phase one has $\rho_1=1/2=\rho_2$. 
In an open TASEP, one obtains a MC phase in the regime where both the entry and exist rates are larger than $1/2$~\cite{blythe}. 
In our model this translates to the following conditions
\begin{subequations}
\begin{eqnarray}
    &&\alpha_1^\text{eff}
    =
    a\frac{N_1}{L}>\frac{1}{2},\,
    \beta_1^\text{eff}
    =
    b \bigg(1-\frac{N_2}{L}\bigg)
    >\frac{1}{2}
    \, , \\
    &&\alpha_2^\text{eff}
    =
    b\frac{N_2}{L}>\frac{1}{2},\,
    \beta_2^\text{eff}
    =
    a\bigg(1-\frac{N_1}{L}\bigg)
    >\frac{1}{2}
    \, ,
\end{eqnarray}
\label{mc-res-no}
\end{subequations}
subject to the overall particle number conservation. 

Given the above relations, we now use the principles discussed in Section~\ref{bound} to determine the various phase boundaries  in Fig.~\ref{phase_diag} for different values of the filling fraction $\mu$. 
To this end, we first consider the boundaries between the phases in which $T_1$ and $T_2$ have spatially uniform densities, belonging to either the LD or the HD phase.
Following our discussion of the phase diagram of an open TASEP  in Section~\ref{bound}, we note that the phase boundary between the LD-LD and HD-HD phases should be given by  $\rho_\text{LD}^{}+\rho_\text{HD}^{}=1$.
Since both $\rho_\text{LD}^{}$ and $\rho_\text{HD}^{}$ are functions of $a,b$ and $\mu$, this condition specifies the boundary in terms of {these parameters}.
Using Eqs.~\eqref{rho-ld-uni} and \eqref{rho_hd}, we find that this condition is satisfied only if $\mu=1$, i.e., coexistence of the LD-LD and HD-HD phases 
is possible only in the half-filled limit. 
Further, noting that ${\rho_\text{LD}^{} \leq 1/2}$, we get by using Eq.~\eqref{rho-ld-uni} along with ${\mu=1}$
\begin{equation}
    a+b>2ab 
\label{coex-boundary}
\end{equation}
as the region of co-existence of the LD-LD and HD-HD phases. 
Since the filling fraction $\mu$ is bounded between 0 and 2, the LD-LD phase is expected to be in the range $0\leq \mu \leq 1$ and the HD-HD phase for $1\leq \mu\leq 2$, which is consistent with the particle-hole symmetry of the model.
Within the co-existence region, the density jump $\rho_\text{HD}^{}-\rho_\text{LD}^{}$, which is the difference in the densities of the HD and LD phases, depends on the position in the phase diagram, i.e., the values of $a$ and $b$. 
Approaching $(a,b) = (1,1)$ on the boundary line $a+b=2ab$ (with $\mu=1$), the density jump is given by 
\begin{equation}
    \rho_\text{HD}^{}-\rho_\text{LD}^{}
    =1-\frac{4ab}{a+b+2ab} 
\end{equation}
vanishes, where we have used Eqs.~(\ref{rho_hd}) and (\ref{rho-ld-uni}), respectively, for $\rho_\text{HD}$ and $\rho_\text{LD}$.

The boundary between the LD-LD and MC-MC phases can be obtained from the condition ${\rho_\text{LD}^{}=1/2}$. 
This gives 
\begin{eqnarray}
  a+b+2ab
  =
  4 a b \mu 
\label{mc-boundary}
\end{eqnarray}
as the boundary parameterized by $\mu$; 
see Fig.~\ref{phase_diag}(b) showing the LD-LD and MC-MC phase boundary for $\mu=5/8$. 
Similarly, the boundary between the HD-HD and MC-MC phases is obtained from the condition $\rho_\text{HD}^{}=1/2$, giving 
\begin{eqnarray}
    a+b+2ab
    =
    4ab\,(2-\mu)
    \, ;
\end{eqnarray}
see Fig.~\ref{phase_diag_hole}(b) showing the HD-HD and MC-MC phase boundary for $\mu=11/8$. 

We now argue that the MC-MC phase can exist only when ${1/2 < \mu < 3/2}$. 
For ${\rho_1=1/2=\rho_2}$, the total particle content of the two TASEP lanes together is $L$. 
Since both $N_1$ and $N_2$ must be non-negative, we must have ${N_0 > L}$. 
This implies that the filling fraction ${\mu = N_0/(2L)}$ must obey the inequality ${\mu > 1/2}$ for the MC-MC phase to exist. 
Likewise, since the carrying capacity of each reservoir is $L$, both $N_1$ and $N_2$ are bounded by $L$. 
Hence, we have ${\mu < 3/2}$ as an upper bound for the MC-MC phase to exist. 
This upper bound can also be obtained by applying the particle-hole symmetry. According to this symmetry, if $\rho$ is a steady state density solution for $\mu$, $1-\rho$ should also be a solution for $2-\mu$; see (\ref{phmu}) and (\ref{ph1}). For the present case, this means if there is no MC--MC phase for $\mu<1/2$, there cannot be any MC--MC phase for $\mu>3/2$.

{ {In this Section~\ref{mean-field theory1}, we have thus used MFT to obtain stationary densities of the uniform phases on the TASEP lanes and the associated reservoir populations. In particular, we obtain the LD phase density $\rho_\text{LD}$, the HD phase density $\rho_\text{HD}$ and the corresponding reservoir populations. 
 In the case of the MC phase on the TASEP lane, our MFT gives only bounds on the reservoir populations, instead of specific values. In the next Section, we use MFT to calculate the nonuniform phases and the transitions between them.}}

\subsection{ Nonuniform lane density: $\rho_1(x)\neq \rho_2(x)$}\label{mean-field theory2}

We now construct the spatially nonuniform mean-field theory solutions with $\rho_1(x)\neq \rho_2(x)$. Nonetheless, $\rho_1(x)$ and $\rho_2(x)$ must still be constructed out of the two basic solutions $\rho_+$ and $\rho_-$, which are given in (\ref{rhopm}) above.   
As discussed earlier, there are two distinct sub-cases:  (a) one or both the lanes have a domain wall each, (b) the two lanes have densities that are spatially uniform but distinct.
 
\subsubsection{LD--DW and LD--HD phases}
 
\textit{LD--DW phase.---} We first focus on the case when there is a single domain wall formed in the bulk of one of the TASEP  lanes.
 For specificity, consider the case where a  DW is formed on lane $T_2$, and lane $T_1$ is in the LD-phase with {uniform} density ${\rho_-(x)=\alpha_1^\text{eff}}$. 
Thus, on lane $T_2$ the DW, positioned at $x_{w2}$, connects a low density regime with density ${\rho_-(x) = \alpha_2^\text{eff}}$  to a high density regime with density  $\rho_+(x)=1-\rho_-(x)$ such that
 \begin{equation}
 \rho_2(x)=\alpha_2^\text{eff} + \Theta(x-x_{w2})(1-2\alpha_2^\text{eff}).\label{rho12}
\end{equation}
 Here $\Theta(x)$ is the Heavyside theta-function defined by ${\Theta(x)=1 (0)}$ for ${x>0(<0)}$. 
We signify this in the following as the  ``LD--DW phase''.


Since the current in a steady state must be uniform and constant, one has
${\alpha_1^\text{eff} (N_1)=\alpha_2^\text{eff}(N_2)=\beta_2^\text{eff} (N_1)}$, resulting in 
\begin{equation}
    f(N_1)=g(N_1)=1-f(N_1)
    \, .
\label{f-g}
\end{equation}
This, in turn, gives
\begin{eqnarray}
    N_1
    =
    \frac{L}{2} \, .
\label{n1-val}
\end{eqnarray}
We then obtain ${\alpha_1^\text{eff} (N_1) = \alpha_2^\text{eff}(N_2)=a/2}$, which in turn gives ${N_2 = La/(2b)}$. 
Thus, \textit{both} $N_1$ and $N_2$ are \textit{independent} of $\mu$. 
We also get, by using the value of $\alpha_1^\text{eff} (N_1)$  
\begin{equation}
    \rho_1
    =
    \frac{aN_1}{L}
    =
    \alpha_1^\text{eff}
    =
    \frac{a}{2}
    \equiv
    \rho_\text{1,LD}^{}
\label{ld-hd-rho1}
\end{equation}
as the bulk density on lane $T_1$.
This further gives 
\begin{subequations}
\label{ld-hd-rho2}
\begin{align}
    \rho_\text{2,LD}
    &= \alpha_2^\text{eff}
    =
    a/2
    =
    \alpha_1^\text{eff}
    \, \\
    \rho_\text{2,HD}
    &=
    1-\beta_2^\text{eff}
    =1-a/2 \, 
\end{align}
\end{subequations}
as the densities of the LD and HD segments of lane $T_2$, respectively. 

For a complete characterization of the domain wall, we need to specify both its position and height.
The position of the domain wall $x_{w2}$ can now be determined by using particle number conservation [see Eq.~(\ref{total-part})] 
\begin{equation}
 N_1 + N_2 + \sum_{j=0}^L \sum_{i=1,2} \rho^i_j
        = 2\mu L.\label{pnc1}
\end{equation}
Using the quasi-continuous coordinate $x$, Eq.~\eqref{pnc1} can be written as
\begin{align}
    \frac{N_1}{L} + 
    \frac{N_2}{L} +
    \int_0^1 dx 
    \big[ 
    \rho_1(x) + \rho_2(x) 
    \big]
    = 2\mu
    \, .
\label{ldwpos}
\end{align}
Using the values of  $\rho_1(x)$ and $\rho_2(x)$ in Eq.~\eqref{ld-hd-rho1} and Eq.~\eqref{ld-hd-rho2} we find
\begin{equation}
    x_{w2}=\frac{(3+a/b)/2-2\mu}{1-a}\, . 
\label{xw2}
\end{equation}
Since the location of the domain wall $x_{w2}, $ is uniquely determined for a given set of parameters $a$, $b$, and $\mu$, the domain wall is \textit{pinned}, i.e., it is a \textit{localized domain wall} (LDW). 

Having obtained the domain wall positions, we now write down the height $\Delta_2$ of the LDW in the LD--DW phase. 
This is given by
\begin{equation}
 \Delta_2\equiv \rho_\text{2,HD} - \rho_\text{2,LD}=1-2\alpha_2^\text{eff}=1-a.\label{DW-height}
\end{equation}


What is the minimum value of $\mu$ required for which the LD--DW phase is possible? 
Particle number conservation can be used to answer this. 
We note that in the LD--DW phase, $N_1=L/2$ [see Eq.~(\ref{n1-val}) above]; 
minimum of $N_2=La/(2b)$ is zero (when $a=0$). The steady-state particle number in $T_2$ depends upon $x_{w2}$. For it to be minimum, $x_{w2}=1$, giving  $2\alpha_1^\text{eff}=a$, which vanishes if $a=0$. This means for the LD--DW phase, $\text{min}(N_0)=L/2$ giving $\mu>1/4$ as the necessary condition for the LD--DW phase. This explains the existence of the LD--DW phase in the phase diagrams with $\mu=3/8,\,5/8,\,7/8$ in Fig.~\ref{phase_diag}. (At $\mu=1$, the LD--DW phase vanishes; see later.)  The mean-field theory profiles of the LDW as above are corroborated by our stochastic simulations studies; see Fig.~\ref{domain-walls} (top). For $\mu>1$, there are HD--DW phases instead of LD--DW phases; see Appendix A.


\textit{LD--HD phase.---} 
In the LD--HD phase, the lanes $T_1$ and $T_2$ are in the LD and HD phases, respectively. 
Thus $\rho_1$ and $\rho_2$ are both uniform but distinct:
We have $\rho_1=\rho_-=1-\rho_2$. 
This means by using Eq.~(\ref{effrate2})
\begin{equation}
    \alpha_1^\text{eff}(N_1)
    =
    \beta_2^\text{eff}(N_1). \label{ld-hd-cond1} 
\end{equation}
Now by using $N_1=L/2$[see  Eq.~\eqref{n1-val} above, we get $\rho_1=a/2$ and $\rho_2=1-\rho_1=1-a/2$ in the LD-HD phase.
%
Since $\rho_1+\rho_2=1$,  the total particle content in $T_1$ and $T_2$ together is just $L$. 
We further get 
\begin{equation}
N_2=N_0-N_1 -L=N_0-3L/2. \label{N2-inhom}
\end{equation}
Since $N_2\geq 0$, we must have $N_0\geq 3L/2$, or $\mu>3/4$ for this phase to exist. Furthermore, since $N_2\leq L$, from Eq.~\eqref{N2-inhom} we must have $N_0\leq 5L/2$ or $\mu<5/4$, in agreement with the particle-hole symmetry of this model. 
Thus, this phase is possible only in the range $3/4 \leq \mu \leq 5/4$. 

\textit{LD--DW and LD--HD phase boundaries.---} 
We now determine the phase boundaries of the LD--DW and LD--HD phases with the other possible phases. In the LD--DW phase, there is a domain wall in $T_2$ at $x_{w2}$, which lies between 0 and 1. When $x_{w2}=1$, the domain wall is shifted to the left boundary of $T_2$, so that $T_2$ is in its LD phase. Therefore, $x_{w2}=1$ gives the condition for the boundary between the LD--DW and LD--LD phases. Similarly, when $x_{w2}=0$ entire $T_2$ is in its HD phase, which then gives the condition for the boundary between the LD--DW and LD--HD phases. 
Setting $x_{w2}=0$ in Eq.~\eqref{xw2} we get the boundary between the LD--DW and LD--HD phases:
\begin{equation}
    a+3b    =     4b\mu \, .
    \label{lhsp1}
\end{equation}
This straight line passes through the origin in the $a-b$ plane. 
The  slope of the line (\ref{lhsp1}) in  the $a-b$ plane is $1/(4\mu -3)$ that approaches infinity as $\mu\rightarrow 3/4$ from above. 
This means as $\mu\rightarrow 3/4$ from above, the line (\ref{lhsp1}) coincides with the $b$ axis. 
Since the LD--HD phase exists for $a+3b<4\mu b$ (i.e., on the left of the line (\ref{lhsp1})), it vanishes as $\mu\rightarrow 3/4$ from above, giving $\mu =3/4$ as the lower threshold of $\mu$ for the existence of the LD--HD phase, a conclusion also obtained above by demanding non-negativity of $N_2$; see discussions below Eq.~\eqref{N2-inhom}.
Just as we find there is a lower threshold on $\mu$ for the existence of the LD--HD phase, there is also an upper threshold on $\mu$ as well.
This upper limit on $\mu$ can be conveniently extracted by using the particle-hole symmetry of the model. We first note that due to the symmetry under interchange of $T_1$ and $T_2$, if $\mu=3/4$ is the lower threshold of $\mu$ for the existence of the LD--HD phase, $\mu=3/4$ must the lower threshold for the existence of the HD--LD phase as well. Now applying the particle-hole symmetry given in Eqs.~(\ref{beq1})-(\ref{beq4}) we conclude that the LD--HD phase, which is the particle-hole symmetric counterpart of the HD--LD phase, can exist only when $\mu<(2-3/4)=5/4$.
We therefore find that the LD--HD phase can exist only in the window $3/4<\mu<5/4$. This is consistent with the fact that there is an LD--HD phase for $\mu=7/8,\,9/8$ but none with $\mu=5/8,\,13/8$ in the phase diagrams in Fig.~\ref{phase_diag} and Fig.~\ref{phase_diag_hole}.

\textit{LD--DW and LD--LD phase boundaries.---} 
We now study the phase boundaries between the LD--DW and LD--LD phases.
When $x_{w2}=1$, both $T_1$ and $T_2$ have the same density that is less than 1/2: $\rho_1=\rho_2= a/2$. 
This marks the boundary of the LD--LD phase with the LD--DW phase. 
Setting $x_{w2}=1$ in (\ref{xw2}), we get the equation for the boundary between the LD--DW and LD-LD phases:
\begin{equation}
    a+b+2ab
    =
    4b\mu
    \, .
\label{llsp1}
\end{equation}
Hence, for $a+b+2ab <4b\mu$, there is an LDW in the bulk of $T_2$, i.e., $x_{w2}<1$.
On the boundary given by Eq.~\eqref{llsp1}, Eq.~\eqref{rho-ld-uni} gives $\rho_\text{LD}=a/2$, showing the continuity of the densities across the boundary (\ref{llsp1}). 

In the parameter regime in the $a-b$ plane that is  bounded between the lines (\ref{lhsp1}) and (\ref{llsp1}) there is a domain wall formed in the bulk of $T_2$, with $T_1$ having a uniform density. In contrast, in the parameter regime $a+3b<4\mu b$, $T_1$ and $T_2$ both have individually uniform bulk (but different) densities, being in their LD and HD phases, respectively.  

We notice that the boundary between the LD--DW and LD--LD phases in the phase diagrams of Fig.~\ref{phase_diag} as given by Eq.~\eqref{llsp1} passes through the origin (0,0) in the $a-b$ plane. 
We have already found that for $\mu<3/4$, the slope of (\ref{lhsp1}), which gives the boundary between the LD--DW and LD--HD phases in the phase diagrams of Fig.~\ref{phase_diag}, turns negative. This makes the LD--HD phase disappear, leaving with only the LD--DW and LD--LD phases. Thus for $\mu<3/4$ the LD--DW phase is possible in the parameter region confined between the $b$-axis and the line given by Eq.~\eqref{llsp1}.  We  then find that the slope of (\ref{llsp1})
\begin{equation}
 \frac{db}{da}=\frac{1}{4\mu-1}
\end{equation}
at the origin, which diverges as $\mu\rightarrow 1/4$. Further at all $a>0$, $db/da<0$ for $\mu<1/4$. Therefore, the parameter region for the LD--DW phase, being  confined between the $b$-axis and the line (\ref{llsp1})  for $\mu<3/4$ must vanish for $\mu<1/4$, indicating disappearance of the LD--DW phase for $\mu<1/4$.
Indeed, we find the LD--DW phase to exist for $\mu=3/8,5/8,7/8$ in the phase diagrams in Fig.~\ref{phase_diag}. Notice that there is no LD--DW phase in the phase diagram for $\mu=1$ in Fig.~\ref{phase_diag}. We return to the phase diagram's structure for $\mu=1$ later.

Lines (\ref{lhsp1}) and (\ref{llsp1}) intersect at $(0,0)$ and $(1,1/(4\mu-3))$; the latter is also their common point of intersection with the line (\ref{mc-boundary}). 
This means that in general, for $b>1/(4\mu-3)$, both the LD--LD and the LD--DW phases cease to exist; instead, one has a transition between the LD-HD and MC-MC phases across the line $a=1$ with $b>1/(4\mu-1)$. See the phase diagram with $\mu=7/8$ in Fig.~\ref{phase_diag}.

Both the LD--HD and LD--DW phases are nonuniform phases, i.e., $\rho_1(x)\neq \rho_2(x)$. However, the role of $\mu$, i.e., effect of changing the total resources $N_0$ on these two nonuniform phases are quite distinct from each other. For example, within the LD--DW phase, which is to be found 
in the phase space region between the boundaries (\ref{lhsp1}) and (\ref{llsp1}),  $N_1=L/2$ [see Eq.~(\ref{n1-val})], a constant, independent of $a,\,b$ and $\mu$, whereas $N_2$ depends explicitly on $a$ and $b$, but not on $\mu$; see Eq.~\eqref{N2-inhom}. This implies that for a given pair of $(a,\,b)$, $\alpha_2^\text{eff}(N_2)=\beta_2^\text{eff}(N_1)$ do not depend upon $\mu$ within the phase space region of the LD--DW phase.
Therefore, within this region, if $\mu$ or $N_0$ varies, the increase or decrease in the particle number is accommodated by an equivalent shift in $x_{w2}$, i.e., by changing the position of the LDW. This ensures that for a given pair of $(a,\,b)$ within the phase space region of the LD--DW phase (i.e., at a given point in the phase diagram), there is {\em no} change in the steady state current $J_\text{LD--DW}$ even when $\mu$ varies.
In contrast, in the LD-HD phase of the model, $N_2$ is independent of $a$ and $b$ since $N_1=L/2$, and the total particles in $T_1$ and $T_2$ are independent of $a$ and $b$. The total particle content in $T_1$ and $T_2$ is just $L$, independent of $\mu$ as well. Hence, any change in $\mu$ must lead to a change in $N_2$ in the LD-HD phase; see (\ref{N2-inhom}) above. This however {\em does not} change the densities in the individual TASEP lanes, since the latter are controlled by $N_1$, which remains fixed in the LD--HD phase, even when $\mu$ changes. Therefore, the steady state bulk current remains unaltered even if $\mu$ varies. We thus see that in the LD--DW phase, a changes in the total resources is reflected in the position of the LDW, whereas in the LD-HD phase it is reflected in $N_2$, the particle content of $R_2$, the reservoir that {\em does not} control the bulk densities in the TASEP lanes in the LD--HD phase.

\subsubsection{DW--LD and HD--LD phases}\label{dw-ld}

In the above analysis for the LD--DW phase, we have assumed a domain wall in $T_2$. 
Instead, a domain wall could be formed in $T_1$, with $T_2$ being in the LD phase, giving the DW-LD phase. Following the logic outlined above 
we find 
\begin{equation}
N_2=L/2, \label{n2-val}
\end{equation}
which is the analog of Eq.~(\ref{n1-val}) and $\rho_2=b/2$. 
Further, the density profile $\rho_1(x)$ now consists of two domains: one with value $b/2$ and the other with value $1-b/2$, meeting at $x_{w1}$ creating a domain wall at that point inside the bulk of $T_1$. Using these values of the two domains in $T_1$ and following the logic outlined above to calculate $x_{w2}$, the position of an LDW in $T_2$ in the LD--DW phase,  we can use the particle number conservation to calculate the location $x_{w1}$ of the LDW in $T_1$ in the DW--LD phase of the system. It is given by
\begin{equation}
 x_{w1}=\frac{(3+b/a)/2-2\mu}{1-b},\label{xw1}
\end{equation}
with a height $\Delta_1=1-b$. The domain wall position $x_{w1}$ lies between 0 and 1. When it is at 1, the LDW in $T_1$ has completely to the right (exit) boundary, leaving entire $T_1$ to be in its LD phase. Thus,
the boundary between the LD-LD phase and DW-LD phase is obtained by setting $x_{w1}=1$. From (\ref{xw1}) we get with $x_{w1}=1$ 
\begin{equation}
 a+b+2ab=4a\mu.\label{llsp2}
\end{equation}
Likewise, when ${x_{w1}= 0}$, the entire $T_1$ lane is in the HD phase with density $\rho_1(x)=\rho_+=1-b/2$, whereas the whole $T_2$ is still in its LD phase, with density $\rho_2(x)=\rho_-=b/2$. This condition yields 
 the analogue of (\ref{lhsp1}) given by
\begin{equation}
 b+3a=4a\mu,\label{lhsp2}
\end{equation}
such that for $b+3a<4a\mu$, the model remains in its HD-LD phase.
The slope of this line is $4\mu -3$, which vanishes as $\mu\rightarrow 3/4$. {The vanishing of the slope as $\mu\rightarrow 3/4$ together with the fact that the line \eqref{lhsp2} passes through the origin (0,0) in the $a-b$ plane implies that the line \eqref{lhsp2} coincides with the $a$-axis as $\mu\rightarrow 3/4$, indicating disappearance of the HD--LD phase. Furthermore,
the lines \eqref{llsp2} and \eqref{lhsp2} intersect at $(1/(4\mu -3),1)$, with this point of intersection approaching $(\infty,1)$ as $\mu\rightarrow 3/4$ from above. 
This is also the point of intersection of (\ref{llsp2}) and (\ref{lhsp2}) with the MC-MC phase boundary (\ref{mc-boundary}).  
Thus, for $\mu\leq 3/4$, the HD-LD phase vanishes, as we have concluded above.} 
For parameter regimes falling
between the lines (\ref{llsp2}) and (\ref{lhsp2}), $T_2$ is in the LD phase, whereas there is a domain wall in the bulk of $T_1$. 
Proceeding similarly as for the LD-HD phase above, we can show that the HD-LD phase can exist only in the range $3/4\leq \mu\leq 5/4$.   
For $a>1/(4\mu-3)$ with $b=1$, there are no DW-LD and LD-LD phases. 
Instead, one has a transition from the HD-LD phase to the MC-MC phase across the line $b=1$. 
Again proceeding similarly to our analysis for the LD--DW phase, we find that the DW-LD phase disappears for $\mu\leq 1/4$.  
These results are connected  with their counterparts for the LD--DW phase through the symmetries of the model.

As in our discussions on the LD--DW and LD--HD phases, both the DW--LD and HD--LD phases are nonuniform, i.e., $\rho_1(x)\neq \rho_2(x)$.  
The effects of varying $\mu$ on the DW--LD and HD--LD phases run exactly parallel to its effect on the LD--DW and LD--HD phases. 
For a given $(a,\,b)$, as $\mu$ changes, the LDW position $x_{w1}$ in $T_1$ changes in the DW--LD phase, leaving the two reservoirs and the TASEP current unchanged. 
In the HD--LD phase, which is controlled by $R_2$ with a population $N_2$, a change in $\mu$ affects the value of $N_1$ leaving everything else unaltered. 

We note that as one considers the limit $a\rightarrow b$, then Eqs.~(\ref{xw2}) and (\ref{xw1}) reveal that 
\begin{eqnarray}
&&x_{w1}\rightarrow 2(1-\mu)/(1-b),\\
&&x_{w2}\rightarrow 2(1-\mu)/(1-a),
\end{eqnarray}
which are identical when $a\rightarrow b$. Furthermore, the domain wall heights $\Delta_1$ and $\Delta_2$ approach each other: $\Delta_2=1-a=1-b=\Delta_1$. Also, if $a=b$ the two TASEP lanes are equivalent, i.e., their stationary density profiles should be statistically identical, suggesting that both the TASEP lanes can have two LDWs with equal height. In this case, the steady-state density is actually a pair of delocalized domain walls; see further discussions and analyses below.

\subsubsection{Stationary currents and steady-state selections}

 We now consider the steady state currents in the phases with $\rho_1(x)\neq \rho_2(x)$. First of all, in the entire inhomogeneous or nonuniform phase regions in the phase diagrams consisting of the LD--HD and LD--DW phases, by using $\rho_1=a/2$ [see Eq.~\eqref{ld-hd-rho1} above], the steady state current is given by
{ \begin{equation}
 J_\text{LD--DW}=\frac{a}{2}\left(1-\frac{a}{2}\right).\label{jsp2}
\end{equation}
Likewise, in the entire inhomogeneous phase consisting of the HD--LD and DW--LD phases, the steady state current is given by
\begin{equation}
 J_\text{DW--LD}=\frac{b}{2}\left(1-\frac{b}{2}\right),\label{jsp1}
\end{equation}}
where we have used $\rho_2=b/2$ [see discussions in Section~\ref{dw-ld} above]. 
 On the other hand, the current in the LD--LD phase is given by
\begin{eqnarray}
 J_\text{LD--LD}&=&\left(\frac{2a\mu}{1+2a+a/b}\right)\left(1-\frac{2a\mu}{1+2a+a/b}\right),
 \label{new-currents}
\end{eqnarray}
where we have used (\ref{rho-ld-uni}).
We now argue that the phase boundaries (\ref{llsp1}) and (\ref{llsp2}) may also be obtained by invoking a minimum current principle~\cite{minm1} by demanding that the currents $J_\text{LD--LD}$ and $J_\text{LD--DW}$ must be equal at the phase boundary. Indeed, 
  the condition $J_\text{LD--LD}=J_\text{LD--DW}$ gives the boundary between the LD--LD and inhomogeneous (LD--HD or LD-DW) phases in the phase diagram: we get 
\begin{equation}
a+b+2ab=4b\mu,
\end{equation}
same as (\ref{llsp1}), such that in the phase space region with $J_\text{LD--LD}<(>)J_\text{DW--LD}$, we get LD--LD (inhomogeneous phases comprising of the LD--DW or LD--HD) phases.
Similarly, the condition $J_\text{LD--LD}=J_\text{DW--LD}$ gives another boundary in the phase diagram between the LD--LD and inhomogeneous (HD--LD or DW--LD) phases as 
\begin{equation}
 a+b+2ab=4a\mu,
 \end{equation}
 that is identical to (\ref{llsp2}), such that in the region with $J_\text{LD--LD}<(>)J_\text{LD--DW}$, we get LD--LD (inhomogeneous, i.e., DW--LD or HD--LD) phases.

 
 \subsubsection{Meeting of the LD--DW and DW-LD phases}
 
 Although our mean-field theory delineates the boundaries of the LD--DW phase with the LD--HD and LD--LD phases, and those of the DW--LD phase with the HD--LD and LD--LD phases,  specifying regions of the LD--DW and DW--LD phases in the $(a-b)$ remains incomplete, as the boundary between LD--DW and DW--LD is yet to be specified. We will discuss this phase boundary below. 
 
 We recall that the phase boundary \eqref{llsp1} separates the phase space regions with the LD--DW phase and the LD--LD phase, whereas the phase boundary \eqref{llsp2} separates the phase space regions with the DW--LD phase and the LD--LD phase.
 These two phase boundaries (\ref{llsp1}) and (\ref{llsp2}) intersect at the origin (0,0) and $(a=2\mu-1,\,b=2\mu-1),\,\mu>1/2$. Since the LD--DW phase exists for $a+b+2ab<4b\mu$ and the DW--LD phase exists for $a+b+2ab<4a\mu$, there is a common region in the $a-b$ plane, where both the LD--DW and DW--LD phases are predicted to exist by the mean-field theory. This raises a question which one among these two is the stable phase.
 We can again invoke the minimum current principle to determine the stable phase in the overlapping region between the LD--DW and DW-LD phases, i.e., between the lines $a+b+2ab<4b\mu$ and $a+b+2ab<4a\mu$.

The steady state currents in DW-LD and LD--DW phases, $J_\text{DW--LD}$ and $J_\text{LD--DW}$, respectively, are given by (\ref{jsp1}) and (\ref{jsp2}), respectively.
From Eqs.~\eqref{jsp2} and \eqref{jsp1}, we find that  $J_\text{DW--LD}<(>)J_\text{LD--DW}$ when $a>(<)b$, with the two currents being equal at $a=b$. Thus,  assuming a minimum current principle,  we conclude that the line $a=b$ forms the boundary between the two domain wall phases. The line $a=b$ meets with the lines (\ref{llsp1}) and (\ref{llsp2}) at $a=b=2\mu-1$, which coincides with the origin (0,0) when $\mu=1/2$. This means the boundary line $a=b$  between LD--DW and DW--LD phases ceases to exist for $\mu\leq 1/2$; also, the lines (\ref{llsp1}) and (\ref{llsp2}) no longer meet each other for $\mu< 1/2$ at any non-zero $(a,b)$; they meet only at the origin (0,0). This means the phase space regions of the LD--DW and DW--LD phases in the $a-b$ plane {\em do not} meet if $\mu<1/2$.   Particle-hole symmetry of the model tells us that for $\mu>1$ (i), the boundary line $a=b$ of the HD--DW and DW--HD phases meets the analogs of (\ref{llsp1}) and (\ref{llsp2}), which are the phase boundaries of the HD--DW and DW--HD phases, respectively, with the HD--HD phase, at $a=b=3-2\mu$, (ii) hence, ceases to exist for $\mu\geq 3/2$, and (iii) these  analogs of the lines (\ref{llsp1}) and (\ref{llsp2}) meet only at the origin for $\mu>3/2$. This in turn means that the DW--HD phase and HD--DW phase {\em do not} meet for $\mu>3/2$.
Incidentally, $\mu=1/2$ and $\mu=3/2$ are also, respectively, the lower and upper thresholds of the existence of the MC--MC phase, as obtained earlier.

\subsubsection{Delocalized domain walls} \label{ddw-dis}

What is the nature of the density profiles on the line $a=b$? Along the boundary $a=b$ between the DW-LD and LD--DW phases (see phase diagrams in Fig.~\ref{phase_diag}) with $\mu=7/8$, $\alpha_1^\text{eff}=\beta_1^\text{eff}=\alpha_2^\text{eff}=\beta_2^\text{eff}$. Thus, there should be two LDWs, one each in $T_1$ and $T_2$. If $x_{w1}$ and $x_{w2}$ are the positions of the two LDWs, the particle number conservation condition gives only a linear relation between them without obtaining them explicitly. This is not surprising; particle number conservation can  be maintained by shifting one LDW by some amount and compensating this by shifting  the second LDW in the reverse
direction. Thus, there are {\em no} unique LDW positions. In fact, the underlying stochasticity of the microscopic dynamics ensures that all possible solutions of $x_{w1}$ and $x_{w2}$ obeying particle number conservation are visited over time. As a result, we actually obtain two long time averaged density profiles, as shown in Fig.~\ref{domain-walls} {(middle)} and {(right)}. These long time averaged $\rho_1(x)$ and
$\rho_2(x)$, unlike an LDW, do not display any discontinuity, but instead take the form of inclined lines, representing the envelopes of the two DDWs. Since the DDWs are to be found only on the boundary between the DW--LD and LD--DW phases (i.e.,$a=b$), which in turn can exist for $1/2<\mu<3/2$, DDWs can be found for $1/2<\mu<3/2$. Outside this window, there are no DDWs.

{

As the long-time average shape of the DDWs is obtained by averaging over the fluctuating domain wall positions, with the extent of the position fluctuations scaling with the system size, mean-field theory cannot predict the shape since the latter ignores all fluctuations. We here provide a set of phenomenological arguments to construct the long-time average shapes of the DDWs. We follow Ref.~\cite{niladri1}. Due to the coexistence of the DW--LD and LD--DW phases on the line $a=b$, we have $N_1=L/2=N_2$; see Eqs.~\eqref{n1-val} and \eqref{n2-val} above. Thus, $N_1+N_2=L$. 
The symmetry of the model generally implies if $a=b$, the two TASEP lanes are equivalent, meaning $\langle \rho_1(x)\rangle = \langle \rho_2(x)\rangle$ for parameters on the line $a=b$, where $\langle..\rangle$ implies temporal averages in the steady states.
Furthermore, notice that the spans of the DDWs may not cover the entire span $L$ of $T_1$ or $T_2$; in Fig.~\ref{domain-walls} (middle), the span covers a fraction of the TASEP lanes with $\mu<1$, whereas in Fig.~\ref{domain-walls}(right), it covers the entire span with $\mu=1$ (half-filled case).  We now estimate the DDW span. We begin by noting that with $\alpha_1^\text{eff}=\beta_1^\text{eff}=\alpha_2^\text{eff}=\beta_2^\text{eff}$, the two DDWs are \textit{statistically identical}, as can be seen in Fig.~\ref{domain-walls} (middle and right) - this holds for all $1/2<\mu<3/2$ for which DDWs are possible. This allows us to write for the mean position of each of the DDWs, $x_0=\langle x_{w1}\rangle = \langle x_{w2}\rangle$. Further note that particles executing asymmetric hopping dynamics in a TASEP necessarily accumulate behind the exit end, i.e., in the present model behind the reservoirs $R_1$ and $R_2$ for TASEP lanes $T_2$ and $T_2$ respectively.
For concreteness, we assume a pair of partial DDWs in $T_1$ and $T_2$ with the remaining parts of $T_1$ and $T_2$ in the LD phases as observed for $\mu<1$ [see, e.g., Fig.~\ref{domain-walls} (middle)]. With $\mu<1$, let us imagine to have replaced each DDW with an LDW at $x_0$ in both $T_1$ and $T_2$, a hypothetical replacement that certainly maintains the particle number conservation and also does not affect the steady-state current. We then apply particle number conservation to get
\begin{equation}
\frac{N_1}{L}+\frac{N_2}{L} + 2\rho_\text{LD}^{} + 2(1-x_0)(1-2\rho_\text{LD}^{})=2\mu.\label{rho-ddw}
\end{equation}
Here, $\rho_\text{LD}^{}=a/2$ and $N_1+N_2=L$. This gives
\begin{equation}
 x_0=-\frac{2\mu-3+a}{2(1-a)}.\label{x-0}
\end{equation}
Thus, at half-filling, i.e., $\mu=1$, $x_0=1/2$, in agreement with our stochastic simulations result; see Fig.~\ref{domain-walls} (right). Since the particles necessarily accumulate immediately behind $R_2$ in $T_1$ and $R_1$ in $T_2$, each DDW wanders a distance ${\cal D}$ measured from the respective exit ends at $x=1$, which is given by
\begin{equation}
 {\cal D}= 1-x_0 \label{ddw-span-1}
\end{equation}
on either side of $x=x_0$. Therefore, the total span of each DDW is $2{\cal D}=2(1-x_0)$. The actual DDW profile may be obtained from the calculated value of ${\cal D}$ in Eq.~\eqref{ddw-span-1} by joining the points $\rho=\rho_\text{HD}=1-\rho_\text{LD}$ at $x=1$ and $\rho=\rho_\text{LD}$ at $x=1-2{\cal D}$. 
Thus, when the system is half-filled ($\mu=1$), we have $x_0=1/2$ from \eqref{x-0} above and hence $2{\cal D}$ covers the entire span of the TASEP lanes; see Fig.~\ref{domain-walls}(right) showing a good agreement between MCS results and our phenomenological construction of the DDW profile. For all other filling factors, $2{\cal D}$ is less than unity, with each DDW spanning a region from $2{\cal D}$ to $x=1$.  In such a case (with $1/2 <\mu<1$), the remaining part between 0 and $2{\cal D}$ of length $1-2{\cal D}$ is in the LD phase. This is  a feature corroborated by our stochastic simulation studies; see Fig.~\ref{domain-walls}(middle) showing a good agreement between MCS results and our phenomenological construction of the partial DDW profile. On symmetry ground, we expect a correspondence between the shapes of the DDWs with $\mu<1$ and $\mu>1$. For $3/2>\mu>1$, DDWs partially span $T_1$ and $T_2$ partially, with the remaining parts of $T_1$ and $T_2$ being in their HD phases. Using the particle-hole symmetry, this is equivalent to partially spanning DDWs with the remaining parts in their LD phases for \textit{holes} with filling fraction $\mu_\text{hole}=2-\mu<1$. The corresponding DDW span can be obtained from Eq.~\eqref{ddw-span-1} by using $\mu_\text{hole}$ as the filling fraction.


\begin{widetext}

\begin{figure}[htb]
\includegraphics[width=5.9cm]{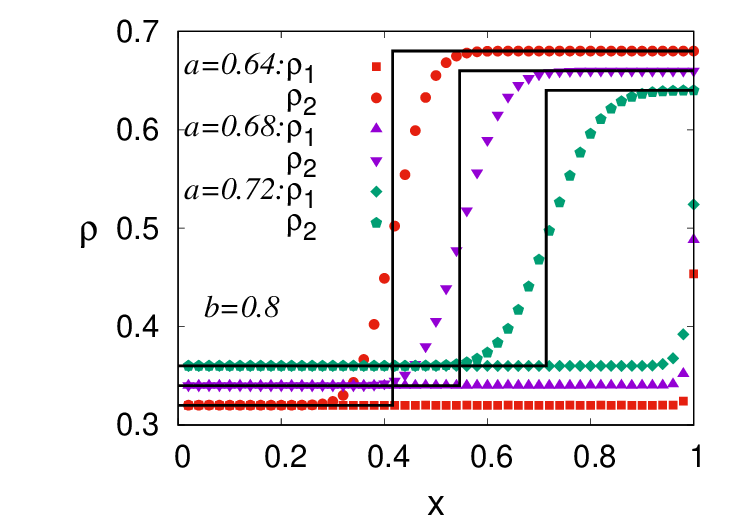}
\includegraphics[width=5.9cm]{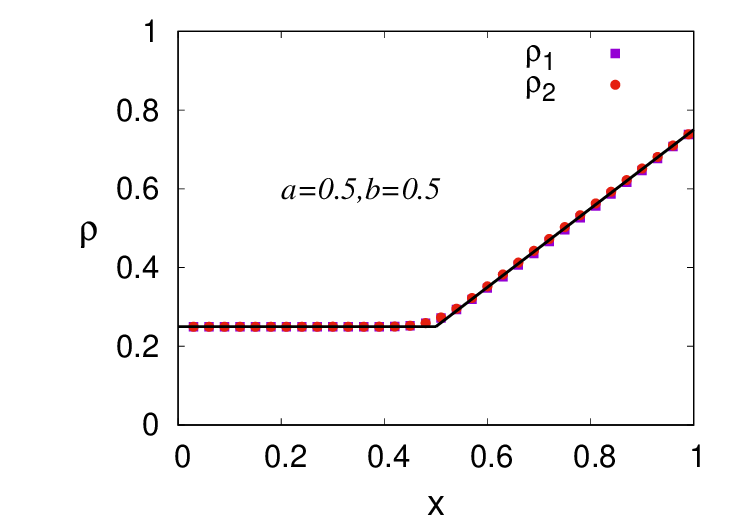}
\includegraphics[width=5.9cm]{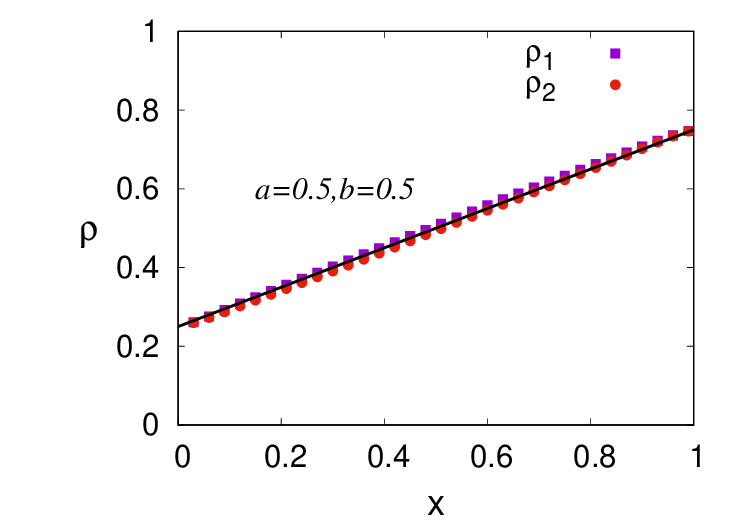}
\caption{
Domain walls in $T_1$ and $T_2$ from stochastic simulations studies. (left) DW--LD: one LDW in $T_1$ and LD phase in $T_2$ with $\mu=7/8$ for various values of $a$; (middle) two DDWs in $T_1$ and $T_2$ with $\mu=7/8$ which partly span the TASEP lanes; (right) two DDWs in $T_1$ and $T_2$ with $\mu=1$ which fully span the TASEP lanes. In each plot, continuous lines correspond to the analytical predictions, and points refer to the stochastic simulations data. System size $L=1000$. 
}
\label{domain-walls}
\end{figure}

 \end{widetext} 

}

\subsubsection{Phase diagram for $\mu=1$}

Let us now construct the phase diagram for $\mu=1$, which is the half-filled limit of the model. When $\mu=1$, lines (\ref{lhsp1}) and (\ref{lhsp2}) are identical and each becomes $a=b$. Thus the LD--DW or DW--LD phases now exist only for parameters on the line $a=b$. Furthermore, the meeting point $(1,1/(4\mu-3))$ of the phase boundaries (\ref{llsp1}) and (\ref{lhsp1}) with (\ref{mc-boundary}) is now at (1,1) when $\mu=1$.  
For parameters on the line $a=b$ between (0,0) and (1,1), one can have one LDW each in $T_1$ and $T_2$. Instead, for the reasons outlined in Sec.~\ref{ddw-dis}, one 
has a pair of DDWs, each of which spans the entire length of $T_1$ and $T_2$, as shown in Fig.~\ref{domain-walls} (right), similar to an isolated open TASEP. The common meeting point of the LD--HD, HD--LD, and MC--MC phases is  (1,1). Thus, the topology of the phase diagram for $\mu=1$ resembles that of an open TASEP; see the phase diagram with $\mu=1$ in Fig.~\ref{phase_diag}. In fact, considering either $T_1$ or $T_2$ only, the phases at $\mu=1$ are those observed in an open TASEP, {\em viz.} LD, HD, and MC phases.

We now give a brief outline of our results on the phases and phase diagrams. The phase diagrams for $\mu=9/8,11/8$ and $13/8$ are given in the Appendix, which may be constructed from those with $\mu=7/8,5/8,3/8$, respectively, by using the particle-hole symmetry of the model. We have found that the topology of the phase diagrams for $\mu=7/8,\,9/8$ hold for the range $3/4<\mu<1$ and $1<\mu<5/4$, respectively, for which the LD-HD and HD-LD phases can exist. For values of $\mu$ outside these ranges, the LD-HD and HD-LD phases disappear, with DW-LD, LD--DW, MC-MC, and LD-LD/HD-HD phases continue to exist for $1/2<\mu<3/4$ and $5/4<\mu<3/2$. For $1/4<\mu<1/2$ or $7/4>\mu>3/2$, DW-LD, LD--DW, and LD-LD or HD-HD phases are the only possible phases. For very low ($\mu<1/4$) and high ($\mu>7/4$) filling, LD-LD and HD-HD phases, respectively, are the only phases possible, with no phase transitions in the $a-b$ plane. For $\mu=1$, i.e., at the half-filled limit, the stable phases are given entirely by the solutions satisfying $\rho_1=1-\rho_2$, with no pure phases like LD-LD or HD-HD phases being 
possible. 
Furthermore, for $1/2 <\mu <3/2$, we find a pair of DDWs existing for parameters on the line $a=b$ between (0,0) and $(2\mu-1,\,2\mu-1)$ for $1/2<\mu <1$, and between (0,0) and $(3-2\mu,\,3-2\mu)$ for  $1<\mu<3/2$. Lastly for $\mu=1$, one gets a phase diagram with LD--HD, HD--LD and MC--MC phases only, with a pair of DDWs for $a,b$ satisfying $a=b$. 
These results on the phases and phase diagrams are validated by our stochastic simulations studies, which agree very well with the corresponding mean-field theory results; see Fig.~\ref{phase_diag}, Fig.~\ref{domain-walls} and Fig.~\ref{phase_diag_hole}. 

{ {To summarize, in this Section~\ref{mean-field theory2}, we have used MFT to calculate the stationary densities in the nonuniform phases. In particular, we have calculated the LD phase density $\rho_\text{LD}$ and HD phase density $\rho_\text{HD}$ in the mixed phases and also the domain wall heights and positions.  Furthermore, we have studied the DDWs in the model, which can be found only for $a,b$, which parametrize the effective entry and exit rates of the TASEP lanes, falling on the line $a=b$ for a range of the filling fraction $\mu$. Lastly, we have shown how a minimum current principle may be used to obtain the phase boundaries between the phases obtained in Section~\ref{mean-field theory1} and Section~\ref{mean-field theory2}, ultimately giving the correct steady states and the full phase diagrams in the $a-b$ plane; see the phase diagrams in Fig.~\ref{phase_diag} and Fig.~\ref{phase_diag_hole}.}}

\section{Nature of the phase transitions}\label{ph-trans}

The phase diagrams in Fig.~(\ref{phase_diag}) have different phases separated by the associated phase boundaries. 
We now discuss the nature of the transitions across these phase boundaries. 
In an open TASEP, the transitions between the LD and HD phases are accompanied by a sudden jump in the bulk density in the TASEP, implying thus a first-order transition with the bulk density acting as the order parameter. 
The density profile corresponding to the parameters on the phase boundary is a DDW that spans the entire open TASEP. 
In contrast, the transitions between the LD or HD and MC phases are second order transitions, with the density difference vanishing smoothly at the respective phase transition. 
In analogy with an open TASEP, the transitions in the present model are described in terms of the densities in each TASEP lane and can be classified in analogy with open TASEPs by noting the density changes. 
Thus in this model, the phase boundaries between the LD--HD and LD--DW, or HD--LD and DW--LD phases are all second-order lines, as the density changes across those phase boundaries are devoid of any discontinuities. Likewise, the phase boundaries between the LD-LD and DW--LD or LD--DW phases are also second order lines for the same reason. 
The phase boundary between DW-LD and LD--DW, the line where a pair of DDW is to be found, is, however, a first order order line, with the $\rho_1$ and $\rho_2$ varying \textit{discontinuously}. 
This is because across this line, $\rho_1(x)$ and $\rho_2(x)$ discontinuously change from being an LDW in the bulk to the LD or HD (depending upon $\mu$) phase.  
Furthermore, the phase boundaries between MC-MC and any other phases are all second-order lines. 
In contrast, the phase boundary between LD--HD and HD--LD (at $\mu=1$) is a first-order line, having the density in \textit{both} the TASEP lanes changing discontinuously.

The above arguments show that the phase diagram with $\mu=7/8$ in Fig.~\ref{phase_diag}  and with $\mu=9/8$ in Fig.~\ref{phase_diag_hole}  have second order phase boundaries each and one first-order boundary each, phase diagrams Fig.~\ref{phase_diag} with $\mu=5/8$ and Fig.~\ref{phase_diag_hole} with $\mu=11/8$ have three second-order boundaries each, phase diagrams with $\mu=3/8$ in Fig.~\ref{phase_diag}  and with $\mu=13/8$ in Fig.~\ref{phase_diag_hole}  have two second-order lines each and phase diagram Fig.~\ref{phase_diag} with $\mu=1$ has two second order and one first-order line. 
In all the phase diagrams, the phase boundaries meet at multicritical points. For example, phase diagram Fig. \ref{phase_diag} with $\mu=1$, $\mu=7/8$, $\mu=5/8$ and $\mu=3/8$ have one, three, one and one multicritical points each. 

A summary of the phases, phase transitions, domain walls, and multicritical points is given in Table~\ref{tab1}.

\begin{widetext}
 
\begin{table}[h!]
 \begin{center}
 \begin{tabular} { |p{3cm}|p{4cm}|p{3cm}|p{3cm}|p{2cm}| }
  \hline
 \multicolumn{5}{|c|}{Steady states of the model} \\
 \hline
 $\mu$ & Phases & Domain walls & Phase boundaries & Multicritical points  \\
 \hline
 $0<\mu<1/4$ & Only LD-LD  & None & None & None \\
 \hline
 $1/4<\mu<1/2$ & DW-LD, LD--DW and LD-LD & One LDW & Two second order & None\\
 \hline
 $1/2<\mu<3/4$ & DW-LD, LD--DW, LD-LD and MC-MC & One LDW & Three second order & None\\
 \hline
 $3/4<\mu<1$ & DW-LD, LD--DW, LD-HD, HD-LD, LD-LD and MC-MC & One LDW and a pair of DDWs& Seven second order and one first order & Three\\
 \hline
 $\mu=1$ & LD-HD, HD-LD and MC-MC & a pair of DDWs & Two second order and one first order & One\\
 \hline
 $1<\mu<5/4$  & DW-HD, HD-DW, LD-HD, HD-LD, HD-HD, and MC-MC & One LDW and a pair of DDW& Seven-second order and one first order & Three\\
 \hline
 $5/4<\mu < 3/2$ & DW-HD, HD-DW, HD-HD and MC-MC & One LDW & Three second order & None\\
 \hline
 $3/2<\mu<7/4$ & DW-HD, HD-DW and HD-HD & One LDW & Two second order & None\\
 \hline
 $7/4<\mu<2$ & Only HD-HD & None & None & None\\
 \hline
 \end{tabular}
 \caption{Table on the summary of the phases, domain walls, phase boundaries, and multicritical points as functions of $\mu$. Rows (6)-(9) with $\mu>1$ can be obtained from the rows (1)-(4) with $\mu<1$ by using the particle-hole symmetry (\ref{phmu})-(\ref{ph3}). } \label{tab1}
 \end{center}

\end{table}
 
\end{widetext}

\section{Particle numbers in the reservoirs}\label{part-num}

We now consider how the steady state values of $N_1,\,N_2$, the populations of the reservoirs $R_1$ and $R_2$, respectively, change across the phase boundaries. 
Consider $\mu<1$. 
Both $\rho_1$ and $\rho_2$ are continuous across the boundaries between the LD--DW phase and  phases like LD--HD and LD--LD phases. 
Since $\rho_1=\alpha_1^\text{eff}(N_1)$, then the continuity of $\rho_1$ across the phase boundaries and linear dependence of $\alpha_1^\text{eff}$ on $N_1$ ensures that $N_1$ must be continuous across those phase boundaries. 
Further, continuity of $\rho_2$ across these phase boundaries and overall particle number conservation demand that $N_2$ must be continuous across those phase boundaries. 
Similar arguments can be constructed to show that both $N_1$ and $N_2$ are continuous across the phase boundaries between DW--LD and phases like HD--LD and LD--LD. 
For $\mu>1$, these results together with particle-hole symmetry of the model discussed in Section~\ref{mft} above can be used to show that both $N_1$ and $N_2$ are continuous across the phase boundaries between DW--HD phase and phases LD--HD and HD--HD phases, and between HD--DW phase and phases HD--LD and HD--HD.  

The transition between the LD-HD and HD-LD phases that occurs on the line $a=b$ with $\mu=1$ is a discontinuous transition for both $T_1$ and $T_2$. 
Now, as we approach the line $a=b$ from either side, both $N_1$ and $N_2$ approach $L/2$, implying that $N_1$ and $N_2$ \textit{do not} show any discontinuity across $a=b$, although the TASEP densities do. This is because as we approach the line $a=b$, $\rho_1+\rho_2=1$, independent of the individual discontinuities of $\rho_1$ and $\rho_2$. 

Since the total particle content in $T_1$ and $T_2$ in the MC--MC phase is just $L$, we must have
\begin{equation}
    N_1+N_2=N_0-L=L(2\mu -1).\label{mc-tot}
\end{equation}
Thus the total of the particles in \textit{both} the reservoirs is a fixed number. 
However, mean-field theory does not set $N_1$ and $N_2$ separately but only specifies their ranges. 
In the MC-MC phase, $\alpha_1^\text{eff}>1/2,\,\alpha_2^\text{eff}>1/2,\,\beta_1^\text{eff}>1/2,\,\beta_2^\text{eff}>1/2$. 
With our choices for the functions $f$ and $g$, these conditions translate into [see Eq.~(\ref{mc-res-no})]
\begin{equation}
    \frac{L}{2a}<N_1< L(1-\frac{1}{2b}),\,\frac{L}{2b}<N_2< L(1-\frac{1}{2a}).
\end{equation}
Thus, $N_1$ and $N_2$ \textit{are not} uniquely determined, but only ranges of their values are found, which are to be supplemented by (\ref{mc-tot}). 
This contrasts with the uniquely determined values of $N_1$ and $N_2$ in all the other phases. 
This further means that continuity of $N_1$ and $N_2$ across the boundaries between MC--MC and any other phases cannot be ascertained.







\section{Summary and outlook}\label{summ}

{
In this study, we investigated how the interplay between the limited availability of particles and the carrying capacity in various regions of a spatially extended system influences the steady-state currents and density profiles within quasi-1D lanes that connect different parts of the system.}{
Specifically, we considered a minimal model consisting of two reservoirs with identical finite carrying capacities, connected by two anti-parallel TASEP lanes of equal length. 
This model in its most general form has five parameters: two entry rates ($\alpha_1$ and $\alpha_2$), two exit rates ($\beta_1$ and $\beta_2$), and the total number of particles, which is conserved. 
Additionally, two functions, $f$ and $g$, govern the effective entry and exit rates, respectively.
For reasons of simplicity, the functions $f$ and $g$ are linked in a simple manner and assumed to have simple forms, which ensure that $f$ is monotonically rising with the reservoir population, whereas $g$ is monotonically decreasing. This ensures that increasing particles in a reservoir leads to increasing flow { to} the TASEP lane from the reservoir, but reduced flow from the TASEP lane to it.} 
The precise quantitative forms of the results depend upon the forms of the functions $f$ and $g$ chosen here. However, even with different functional forms for $f$ and $g$, as long as they are monotonically increasing and decreasing with their arguments, together with finite resources and carrying capacities, we expect the qualitative nature of the results here to hold. 


We focus on the case $\alpha_1=\beta_2=a,\,\alpha_2=\beta_1=b$. {To study the model, we have used a complementary approach combining mean-field theory with extensive stochastic simulations. The mean-field theory predictions agree very well with the stochastic simulations results both qualitatively and quantitatively.}

{The model reveals surprising phase behavior that deviates from previous studies of TASEP models with finite resources~\cite{reser1,reser2,reser3,brackley,astik-1tasep,arvind1,arvind2,arvind3,sourav1,sourav2}.
First and foremost, we find {the two TASEP lanes can exhibit the same or different phases depending on the parameters. This includes scenarios such as having a single LDW in one lane or a pair of DDWs in both lanes.} 
{The ensuing phase diagram is fairly complex, particularly at moderate system filling; for example, at $\mu = \frac{7}{8}$, there are six distinct phases. A distinguishing feather of  phase diagram is that while its topology is very different from the phase diagram of an open TASEP in general, at $\mu=1$, it resembles that of an open TASEP for each of the TASEP lanes $T_1$ and $T_2$.  } 
The associated phase transitions and the boundaries are generally second-order in nature, although for $\mu=1$, a first-order transition appears. We are able to calculate all of the phase boundaries analytically within mean-field theory.
We can also identify the different threshold values of $\mu$ at which different transitions occur. 
Furthermore, we calculate the occupations $N_1$ and $N_2$ of the two reservoirs. We find that they are generally unequal. In fact, the populations of the two reservoirs could be preferentially controlled, i.e., getting them relatively populated or depopulated, by appropriate choice of the control parameters. This can lead to steady population imbalances and highly inhomogeneous particle distributions between different parts of the systems. 


{
We have here investigated the simplest scenario involving two reservoirs connected by two anti parallel TASEP lanes with finite availability and finite carrying capacity of resources. 
This framework can be easily extended to any number of reservoirs and TASEP lanes. 
While our mean-field theory can be adapted to provide specific quantitative predictions for such extended systems, we expect the qualitative features observed in our study to persist. 
For example, when the filling factor $\mu$ is very small or very large, all TASEP lanes should be in their LD or HD phases, respectively. 
As $\mu$ increases, mixed phases of the TASEP lanes are anticipated to appear.}
Further, and as a consequence of the global particle number conservation, independent of the numbers of the reservoirs and TASEP lanes, such a generalized model could display only one LDW or DDWs of any number more than one at some points in the phase space.  
Similar to the present study, we expect that the individual reservoir populations in the steady states can be controlled by the TASEP lanes together with appropriate choice for the model parameters. The quantitative extent of these population distributions should depend upon the precise interconnectivity of the reservoirs by the TASEP lanes. One can also consider using functions defining the effective entry and exit rates with some having very different functional form than the others; e.g., one or some of them could of the kind used here and the remaining ones could be similar to that studied in Ref.~\cite{astik-1tasep}. Such a ``mixed'' model will allow us to study possible competition between rate functions of different types of functions, leading to complex phase behavior. Detailed stochastic simulations studies should be able to validate these qualitative predictions. We hope our studies here will provide impetus to research along  these directions.



\begin{acknowledgments}
AB thanks the SERB, DST (India) for partial financial support through the CRG scheme [file no.: CRG/2021/001875]. 
The work of EF was funded by the Excellence Cluster ORIGINS under Germany’s Excellence Strategy -- EXC-2094 -- 390783311. AH thanks the Alexander von Humboldt Stiftung, Germany for a postdoctoral fellowship. AB and EF thank the Alexander von Humboldt Stiftung, Germany for financial support within its Research Group Linkage Programme framework (2024). 
\end{acknowledgments}


\appendix


\section{Phase diagrams for $\mu>1$}\label{phase-great}

In this Section, we present our mean-field theory and stochastic simulations results for the phase diagrams with $\mu>1$. In  Fig.~\ref{phase_diag_hole}, we show the phase diagrams for $\mu=9/8, 11/8, 13/8$.  The mean-field theories for the phase diagrams in Fig.~\ref{phase_diag_hole} can be obtained from those in Fig.~\ref{phase_diag} by applying the particle-hole symmetry (\ref{phmu})-(\ref{ph3}), giving the phases with $\mu=9/8,11/8,13/8$, respectively, in terms of the phases with $\mu=7/8,5/8,3/8$.  Thus the phase diagrams in Fig.~\ref{phase_diag} and Fig.~\ref{phase_diag_hole} together validate the  particle-hole symmetry (\ref{phmu})-(\ref{ph3}) set up above, which in turn are in agreement with the corresponding stochastic simulations results.

\section{Simulation algorithm}
\label{sim-algo}


In our stochastic simulations, we have used random-sequential updating~\cite{rand1} of the sites of the TASEP and reservoirs, subject to the update rules (a)-(c) described
above in Sec.~\ref{model}. 
In each iteration, we random-sequentially choose one of the reservoirs or a site from each TASEP lane for being updated. 
The particles enter into $T_1$ or $T_2$ at the $j=1$ and $j=L$ site, respectively, from the specific reservoir attached to them, subject to exclusion, at rates $\alpha_{1,2}^\text{eff}(N_{1,2})$ which is dynamically calculated at every stochastic simulations step. 
After hopping through the system from $j = 1$ to $L$ in $T_1$ and $j=L$ to 1 in $T_2$, subject to exclusion, in each of $T_1$ and $T_2$, the particles exit the
system from $j = L$ of $T_1$ and $j=1$ of $T_2$ at dynamically calculated rates $\beta_{1,2}^\text{eff}(N_{2,1})$, and enter the reservoir attached to the $i=L$ site. 
The density profiles are calculated after reaching the steady states, and temporal averages are performed. 
This produces the time-averaged, space-dependent density profiles.

\begin{figure*}[!t]
\centering
\includegraphics[width=8.8cm]{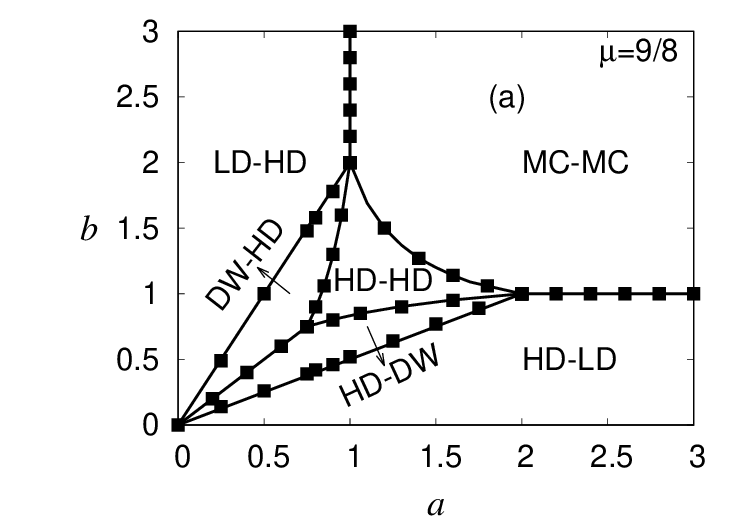}\hfill\includegraphics[width=8.8cm]{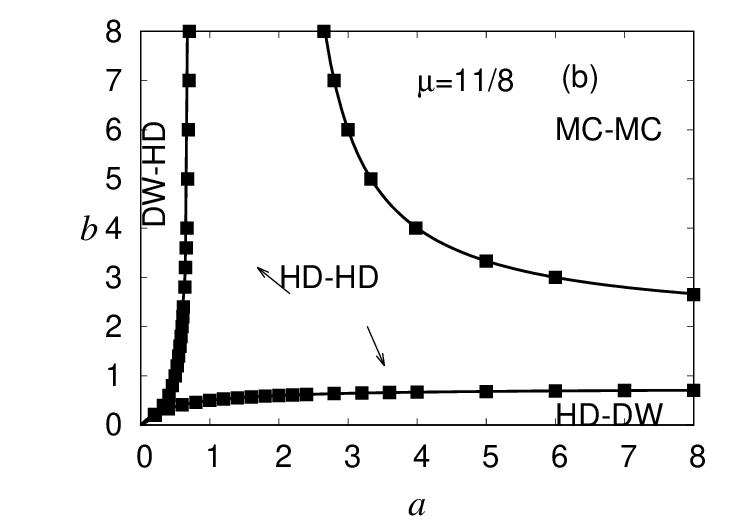}
\\
\includegraphics[width=8.8cm]{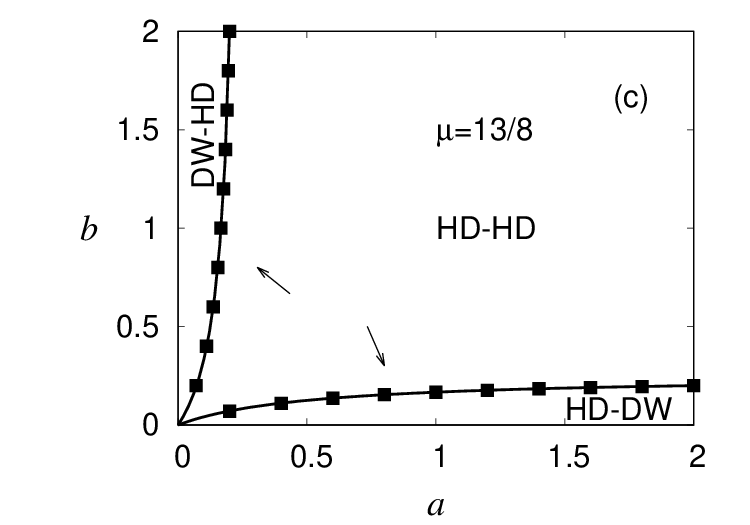}
\caption{
(a) Phase diagrams in the $a-b$ plane obtained from mean-field theory (continuous lines) and stochastic simulations (discrete points) studies. (a) $\mu=9/8$: HD--HD, LD--HD, HD--LD, DW--HD, HD--DW and MC--MC. This phase diagram can be obtained from the
phase diagram in Fig.~\ref{phase_diag} with $\mu=7/8$ in the main text by using the particle-hole symmetry (see text) and is valid for $1<\mu<5/4$. (b) $\mu=11/8$: The phases are DW--HD, HD--DW, HD--HD and MC--MC.  This phase diagram can be obtained from the phase diagram in Fig.~\ref{phase_diag} with $\mu=5/8$ in the main text by using the particle-hole symmetry (see text),(c)  $\mu=13/8$: The phases are DW--HD, HD--DW and HD--HD. 
 This phase diagram can be obtained from the phase diagram in Fig.~\ref{phase_diag} with $\mu=3/8$ in the main text by using the particle-hole symmetry (see text). These demonstrate the particle-hole symmetry of the model.}
\label{phase_diag_hole}
\end{figure*}

\bibliography{2Tasep-ref.bib}

\end{document}